\documentclass[
reprint,
amsmath,
amssymb,
aps,
superscriptaddress,
nofootinbib
preprintnumbers,
prd
]{revtex4-2}
\usepackage{hyperref}
\hypersetup{colorlinks=true, citecolor=blue, urlcolor=blue, linkcolor=blue}
\usepackage[T1]{fontenc}
\DeclareRobustCommand{\VAN}[3]{#2}
\let\VANthebibliography\thebibliography
\def\thebibliography{\DeclareRobustCommand{\VAN}[3]{##3}\VANthebibliography}

\def\thebibliography{\DeclareRobustCommand{\VAN}[3]{##3}\VANthebibliography}
\usepackage{graphicx}	
\usepackage{amsmath}	

\usepackage{amssymb}	
\usepackage{bm}		

\usepackage{ae,aecompl}
\usepackage{newtxtext,newtxmath}
\usepackage{xurl}

\usepackage{dcolumn}    		
\usepackage{aas_macros}
\usepackage{tipa} 
\usepackage[table]{xcolor}

\renewcommand{\selectlanguage}[1]{} 

\date{\today}

\begin{document}

\title[Neural Network Emulators: Assessing Tension Metrics]{Attention-Based Neural Network Emulators for Multi-Probe Data Vectors Part II: Assessing Tension Metrics}

\author{
Evan Saraivanov
}
\email{
evan.saraivanov@stonybrook.edu
}
\affiliation{Department of Physics and Astronomy, Stony Brook University, Stony Brook, NY 11794, USA }
\author{
Kunhao Zhong
}
\affiliation{Department of Physics and Astronomy, Stony Brook University, Stony Brook, NY 11794, USA }
\affiliation{Department of Physics and Astronomy, University of Pennsylvania, Philadelphia, PA 19104, USA }
\author{
Vivian Miranda
}
\affiliation{C. N. Yang Institute for Theoretical Physics, Stony Brook University, Stony Brook, NY 11794, USA }
\author{
Supranta S. Boruah
}
\affiliation{Department of Astronomy and Steward Observatory, University of Arizona, 933 N Cherry Ave, Tucson, AZ 85719, USA }
\affiliation{Department of Physics and Astronomy, University of Pennsylvania, Philadelphia, PA 19104, USA }
\author{
Tim Eifler
}
\affiliation{Department of Astronomy and Steward Observatory, University of Arizona, 933 N Cherry Ave, Tucson, AZ 85719, USA }
\affiliation{Department of Physics, University of Arizona,  1118 E Fourth Str, Tucson, AZ, 85721-0065, USA }
\author{
Elisabeth Krause
}
\affiliation{Department of Astronomy and Steward Observatory, University of Arizona, 933 N Cherry Ave, Tucson, AZ 85719, USA }
\affiliation{Department of Physics, University of Arizona,  1118 E Fourth Str, Tucson, AZ, 85721-0065, USA }

\begin{abstract}
{
The next generation of cosmological surveys is expected to generate unprecedented high-quality data, consequently increasing the already substantial computational costs of Bayesian statistical methods. This will pose a significant challenge to analyzing theoretical models of cosmology. Additionally, new mitigation techniques of baryonic effects, intrinsic alignment, and other systematic effects will inevitably introduce more parameters, slowing down the convergence of Bayesian analyses. In this scenario, machine-learning-based accelerators are a promising solution, capable of reducing the computational costs and execution time of such tools by order of thousands. Yet, they have not been able to provide accurate predictions over the wide prior ranges in parameter space adopted by Stage III/IV collaborations in studies employing real-space two-point correlation functions. This paper offers a leap in this direction by carefully investigating the modern transformer-based neural network (NN) architectures in realistic simulated Rubin Observatory year one cosmic shear $\Lambda$CDM inferences. Building on the framework introduced in Part I, we generalize the transformer block and incorporate additional layer types to develop a more versatile architecture. We present a scalable method to efficiently generate an extensive training dataset that significantly exceeds the scope of prior volumes considered in Part I, while still meeting strict accuracy standards. Through our meticulous architecture comparison and comprehensive hyperparameter optimization, we establish that the attention-based architecture performs an order of magnitude better in accuracy than widely adopted NN designs. Finally, we test and apply our emulators to calibrate tension metrics.
}
\end{abstract}

\maketitle

\section{Introduction}

With the approach of the next generation of cosmological surveys, an unprecedented amount of high-quality data will become available to investigate of the nature of late-time dark energy. These experiments will measure the Cosmic Microwave Background~\citep{Planck:2018nkj,ACT:2020gnv,SPT-3G:2021eoc}, type Ia supernovae~\citep{Pan-STARRS1:2017jku,2022ApJ...938..113S}, Baryon Acoustic Oscillations~\citep{Ross:2014qpa, BOSS:2016wmc, Raichoor:2020vio, eBOSS:2020yzd, 2022MNRAS.511.5492Z}, and the lensing and clustering of optical galaxies~\citep{HSC:2018mrq,Hamana:2019etx, LSST:2019wqx, KiDS:2020suj, Heymans:2020gsg, DES:2021wwk}. The expected enhancement in experimental precision poses a new challenge to theoretical predictions; they will become more complicated and computationally expensive. For instance, new effective field theory methods provide physically motivated nuisance parameters in galaxy clustering surveys while adding several new parameters that must be sampled with Markov Chain Monte Carlo methods (MCMC)~\citep{Kokron:2021xgh, Mergulhao:2021kip, Bakx:2023mld, Chen:2023yyb, Nicola:2023hsd, Rubira:2023vzw}. 

At the same time, the past twenty years has seen a stagnation in the single-core performance of computers \footnote{\url{https://github.com/karlrupp/microprocessor-trend-data}}. The famous Moore Law that, which states that the number of transistors on an integrated circuit doubles every two years, has only been upheld by introducing processors with many dozens of CPU cores. This single-core stagnation poses a problem as most cosmology codes have limitations in their shared memory parallelization via OpenMP~\citep{dagum1998openmp}) that prevent them from being scaled by more than $\mathcal{O}(10)$ threads. These widely adopted codes in statistical inferences include the Boltzmann codes CLASS~\citep{CLASS-code} and CAMB~\citep{Lewis:2002ah, Howlett_2012}. 

A promising approach to alleviate this problem is the adoption of neural networks or other machine learning algorithms to accelerate some portion of the data vector computation~\citep {Auld:2006pm, Auld:2007qz, Albers:2019rzt, Arico:2021izc, 2022MNRAS.511.1771S, nygaard_connect_2022,bonici2022fast, Arico__2022, Donald_McCann_2022, G_nther_2022,10.1093/mnras/staa2102}. Current emulators often emulate data in harmonic space; for instance, the \textsc{CosmoPower} mimics various power spectra using neural networks~\citep{2022MNRAS.511.1771S}, and the \textsc{Euclid Emulator v2.0} simulates the nonlinear matter power spectrum using polynomial chaos expansion~\citep{Euclid:2020rfv}. However, two-point correlation functions in real space are closer to what is being measured on galaxy catalogs and are widely adopted in current and upcoming collaborations. These include the Dark Energy Survey~\citep{DES:2016jjg, DES:2017tss, DES:2017qwj, DES:2020ahh, 2021PhRvL.126n1301T, DES:2021zxv, DES:2021rex, DES:2022ygi}, KiDS~\citep{KiDS:2020suj}, HSC~\citep{Hamana:2019etx} and the Rubin Observatory~\citep{DESC-SRD}. 

In our recent work,~\citet{zhong2024attentionbased} started the exploration of the novel \textit{transformer} model equipped with \textit{scaled dot product attention}, and represented a significant leap. We trained our emulators with three training sets containing two, four, and eight million models selected from a mixture of Latin Hypercube and uniform samplings. 
The prior range for the cosmological parameters was similar to the one adopted by the \textsc{Euclid Emulator v2.0} for the matter power spectrum~\citep{Euclid:2020rfv}. We then applied the emulators to forecast the consistency between growth and geometry parameters in the growth-geometry split models~\citep{Wang:2007fsa, 2015PhRvD..91f3009R, DES:2020iqt, Ruiz-Zapatero:2021rzl, Nguyen:2023fip, Zhong:2023how}. 

However, the prior adopted in~\citet{zhong2024attentionbased} is still informative in statistical inferences that simulate the capabilities of the Dark Energy Survey and Rubin Observatory, typical examples of stage III/IV surveys~\citep{Albrecht:2006um}. Another limitation of our initial study is the somewhat restricted scalability, with respect to the number of free parameters, of the uniform sampling used to select the training points. Models incorporating dozens of additional parameters may require tens of millions of training points to be emulated accurately. Additionally, the neural network in~\citet{zhong2024attentionbased} consisted only of transformers and did not consider the combination of different architectures. 

This is the second of three manuscripts devoted to the attention-based architecture used commonly by large language models~\citep{vaswani_attention_2017}. Our analysis describes network designs, training and validation procedures, and choices of hyperparameters that enable the emulation of cosmic shear over a large volume in parameter space. Our findings, here restricted to cosmic shear, can be generalized to accelerate the modeling of galaxy-galaxy lensing, galaxy clustering, cluster lensing and clustering~\citep{DES:2020uce, 2021PhRvL.126n1301T}, and all cross-correlations between galaxy shapes, positions, and CMB observables~\citep{Xu:2023qmp, Fang:2023efj}. 

This work expands the investigation in~\citet{zhong2024attentionbased}, building training samples valid on priors that, albeit not broad enough to cover the entire parameter range adopted in recent weak lensing studies, are so comprehensive that they will allow follow-up studies of~\citet{DES:2020hen} and~\citet{2023MNRAS.521.1184L} to be completed with reasonable computational resources. We then exemplify these potential applications by comparing the accuracy of \textsc{PolyChord}~\citep{2015MNRAS.450L..61H} and \textsc{Nautilus}~\citep{Lange:2023ydq} in the computation of Bayesian evidence. 

In~\citet{zhong2024attentionbased}, we design machine learning emulators with pure architectures; for example, the transformer-based emulator only contained transformer blocks. Here, we generalize these architectures, and our final design has mixed types of building blocks. We also expand the definition of the transformer blocks, allowing the parallel dense layers that follow the self-attention mechanism to have independent trainable parameters (they are not all identical, as in~\citet{vaswani_attention_2017}).

Our approach is orthogonal to the recent studies such as~\citet{Boruah:2022uac} and~\citet{2023JCAP...01..016T}, as they aim to emulate the $68\%$ and $95\%$ confidence regions in posteriors produced by MCMCs in a small volume in parameter space. Their training is heavily concentrated on samples near a single fiducial cosmology, focusing on their emulators being computationally inexpensive to train. Specifically,~\citet{Boruah:2022uac} and ~\citet{2023JCAP...01..016T} require less than one hundred thousand points to train their networks, and they use MCMCs to generate the training data. On the other hand, large changes in the adopted fiducial cosmology, which can happen when forecasting or selecting a new combination of datasets that shift the high-likelihood region, require retraining.

The limited volume of applicability in parameter space of past neural networks has prevented emulators from supporting two computationally intensive investigations explored by the Dark Energy Survey: the assessment of tension metrics~\citep{DES:2020hen} and the comparison of samplers~\citep{2023MNRAS.521.1184L}. Emulators with sufficient training coverage can accelerate the calculation of Bayesian evidence~\citep{2022arXiv220711457B, Hu:2023bzz} metrics. The LSST Dark Energy Science Collaboration (DESC) will have to calibrate the different tension metrics they plan to adopt when analyzing the upcoming Rubin data. However, the more stringent accuracy requirements of Stage-IV surveys will further increase the formidable computational costs associated with these calculations. This is where our emulator can help.

We show that the combination of our proposed emulator and the optimized \textsc{Nautilus} sampler allows Bayesian evidence to be computed using a single CPU core in a few hours. For the first time, MCMCs and Bayesian evidence in weak-lensing inferences do not require supercomputers to be evaluated. We then perform two significant shifts in the data vector and compare what a few tension metrics predict to the tension between LSST-Y1 and Planck, loosely following~\citet{DES:2020hen}. However, we add a few hundred noise realizations in the LSST-Y1 data vectors and investigate how they affect these metrics.

\section{Neural Network Emulator}

\subsection{Architectures}

\begin{figure}
    \centering
    \includegraphics[width=\columnwidth]{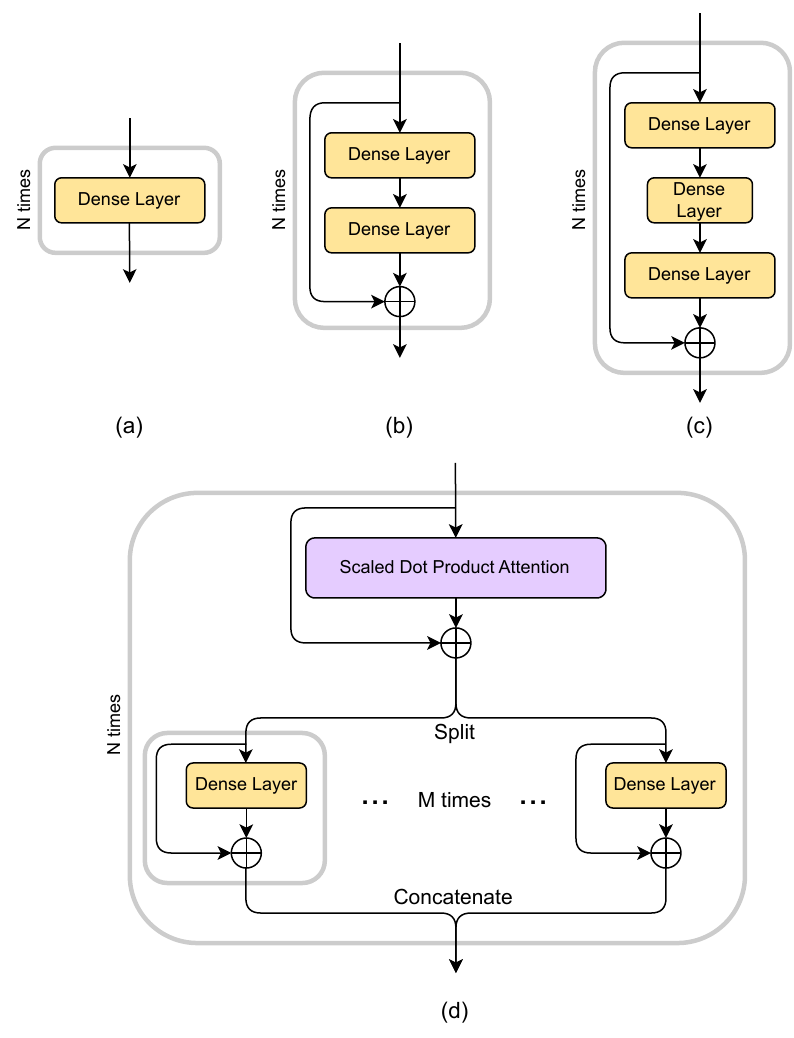}
    \caption{A summary of the pure architectures considered in this manuscript: they all consist of a building block repeated $N$ times. \textit{Top left}: A multi-layer perceptron (MLP) constructed as a sequence of dense layers. \textit{Top middle}: A residual network (\textsc{ResMLP}), consisting of a sequence of two dense layers with connections allowing information to skip them. \textit{Top right}: A bottlenecked \textsc{ResMLP}, where each building block has an additional embedding layer of a smaller dimension. \textit{Bottom}: A pure transformer design consisting of {\it N} transformer blocks, each being a scaled dot-product attention layer and a sequence of MLPs. The input vector must have a size that is multiple of {\it M}, as we break it into {\it M} channels to perform the attention operation. Each channel is then sent to its own dense layer with a skip connection. Finally, one can concatenate these designs to construct mixed architectures, such as the one we promote in Fig.~\ref{fig:full_arch} that combines \textsc{ResMLP} and transformers. Our adopted transformer implementation differs from~\protect\cite{vaswani_attention_2017} and~\protect\cite{zhong2024attentionbased} because the {\it M} parallel dense layers are not identical. Here, and throughout this work, we deliberately simplify notation by calling the building blocks the same name as the underlying architecture. For instance, a single \textsc{ResMLP} block would be equivalent to the building block of the \textsc{ResMLP} architecture.}
\label{fig:arches}
\end{figure}

In this section, we examine four architectures for constructing feed-forward neural networks, the first being the multi-layer perceptron (MLP). Sequential, feed-forward neural networks are directed graphs that can be partitioned into layers based on their distance from the input nodes. MLP is a series of dense (or fully connected) layers, shown on panel (a) in Fig.~\ref{fig:arches}. Each fully connected layer consists of a $\tilde{d}\times d$ matrix and a $\tilde{d}$ dimensional vector, followed by an activation function. Unless explicitly stated otherwise, we assume the input and output dimensions to be the same, $\tilde{d}=d$.

\begin{figure}
    \centering
    \includegraphics[width=0.9\columnwidth]{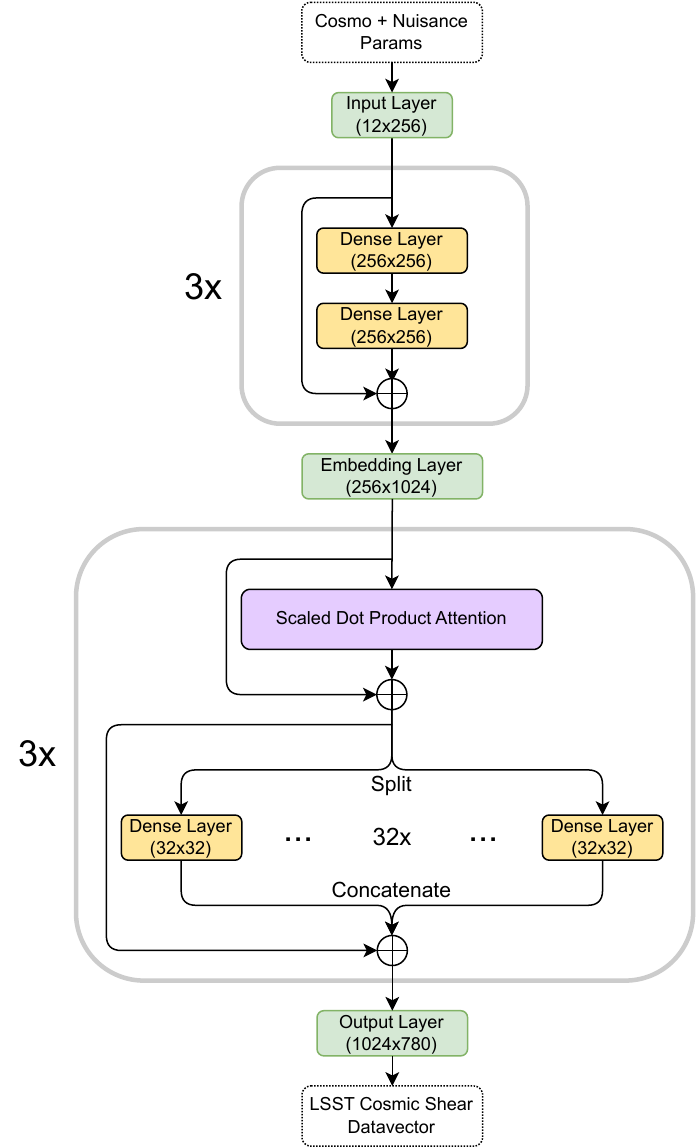}
    \caption{The complete attention-based architecture, denoted as \textsc{ResTRF}, adopted throughout this manuscript. The input layer transforms the cosmological and nuisance parameters into a larger vector of size 256. The \textsc{ResTRF} architecture starts with three residual blocks with width 256, which performed well up to the $T=64$ training temperature without any additional transformer blocks. We then append one transformer block at the end to push the emulator to higher training temperatures. The embedding layer increases the dimensionality of the \textsc{ResMLP} output from 256 to 1024, allowing us to break it into 32 channels of size 32. We then perform the attention operation, and each channel is sent to an independent dense layer (we also included a skip connection). We repeat the transformer block two more times, after which we concatenate the channels so the output data vector also has a size of 1024. Finally, the output layer reduces the data vector size to match the number of data points in the LSST cosmic shear data vector. The input and output layers are linear transformations without additional nonlinear activation functions.}
    \label{fig:full_arch}
\end{figure}

\begin{table}
    \centering
    \renewcommand{\arraystretch}{1.3}
    \begin{tabular}{lcr}
        \hline
        Parameter & Value & Prior \\\hline
        \textbf{Standard Cosmology}\\
        $\log(10^{10} A_{\rm s})$ & $3.047$   & $\mathcal{U}(1.61,3.91)$  \\
        $n_{\rm s}$                & $0.9665$  & $\mathcal{U}(0.87,1.07)$  \\
        $h \equiv H_0/100$                & $0.6766$   & $\mathcal{U}(.55,.91)$      \\
        $\omega_\mathrm{b}$    & $0.02242$ & $\mathcal{U}(0.01,0.04)$  \\
        $\omega_\mathrm{c}$    & $0.11933$ & $\mathcal{U}(0.001,0.99)$ \\\hline
        \textbf{Source photo-z} \\
        $\Delta z_{\mathrm{s},1}$     & $0$       & $\mathcal{N}(0,0.005)$    \\
        $\Delta z_{\mathrm{s},2}$     & $0$       & $\mathcal{N}(0,0.002)$    \\
        $\Delta z_{\mathrm{s},3}$     & $0$       & $\mathcal{N}(0,0.002)$    \\
        $\Delta z_{\mathrm{s},4}$     & $0$       & $\mathcal{N}(0,0.003)$    \\
        $\Delta z_{\mathrm{s},5}$     & $0$       & $\mathcal{N}(0,0.002)$    \\\hline
        \textbf{Intrinsic Alignment} \\
        $a_1$  & $0.5$     & $\mathcal{U}(-5,5)$       \\
        $\eta_1$ & $0$       & $\mathcal{U}(-5,5)$       \\\hline
        \textbf{Shear calibration} \\
        $m_1$                & $0$       & $\mathcal{N}(0,0.005)$    \\
        $m_2$                & $0$       & $\mathcal{N}(0,0.005)$    \\
        $m_3$                & $0$       & $\mathcal{N}(0,0.005)$    \\
        $m_4$                & $0$       & $\mathcal{N}(0,0.005)$    \\
        $m_5$                & $0$       & $\mathcal{N}(0,0.005)$    \\\hline
        \textbf{Cosmology 2} \\
        $\log(10^{10} A_{\rm s})$ & $3.013$   & \\
        $\Omega_\mathrm{m}$       & $0.336$   & \\\hline
        \textbf{Cosmology 3} \\
        $\log(10^{10} A_{\rm s})$ & $3.109$   & \\
        $\Omega_\mathrm{m}$       & $0.284$   & \\\hline
        \textbf{Cosmology 4} \\
        $\log(10^{10} A_{\rm s})$ & $3.041$   & \\
        $\Omega_\mathrm{m}$       & $0.317$   & \\\hline
        \textbf{Cosmology 5} \\
        $\log(10^{10} A_{\rm s})$ & $3.065$   & \\
        $\Omega_\mathrm{m}$       & $0.304$   & \\\hline
    \end{tabular}
    \caption{
    The fiducial cosmology adopted when training the Neural Network and assessing tension metrics. We choose to center the LSST cosmology at the Planck 2018 best fit~\citep{noauthor_planck_2018, Planck:2018vyg} to facilitate the application of calibrating tension metrics.
    }
    \label{tab:cosmology}
\end{table}

MLPs are commonly used in cosmological emulators and have proven sufficient for certain applications~\citep{2022MNRAS.511.1771S}. However, MLPs have some drawbacks compared to more modern architectures, the most prominent being that information from the first layers is masked by gradients in the deep layers. Thus, the gradient is prone to vanishing exponentially~\citep{279181,pascanu2013difficulty}. To alleviate this, one can add residual connections, which allow the gradient from shallow layers to propagate directly to the deeper layers~\citep{he_deep_2015}. We refer to this architecture as a \textsc{ResMLP}, shown on panel (b) in Fig.~\ref{fig:arches}.
 
Despite alleviating the vanishing gradient problem, nodes in a \textsc{ResMLP} may carry information that does not contribute to the output, making training more difficult~\citep{he_deep_2015}. To resolve this, one can force the data into an embedding dimension $\tilde{d} < d$, where the unnecessary information can be forgotten. Then, the network can expand back to the original dimension $d$ before adding the residual connection. We refer to this model as a \textsc{ResBottle}, illustrated on panel (c) in Fig.~\ref{fig:arches}.

Attention is a modern neural architecture that learns sequential data with long-range and long-term interactions~\citep{vaswani_attention_2017}. This model examines the {\it similarity} between input vectors; the mathematical equation defining this operation distinguishes distinct types of attention. To define the one adopted by our investigation, suppose the network is given a sequence of $M$ input vectors $X=\{\mathbf{x}_1,\mathbf{x}_2,\ldots,\mathbf{x}_M\}$. The dot products of each $x_{i=1,\ldots,M}$ can be used as coefficients to define a new set $Z=\{\mathbf{z}_1,\mathbf{z}_2,\ldots,\mathbf{z}_M\}$ of $M$ output vectors with
\begin{equation}\label{eqn:dot-products}
	\mathbf{z}_k \equiv W \, \mathbf{x}_k \, \,
\end{equation}
with the elements $w_{i,j}$ of matrix $W$ being $w_{i,j} \equiv \mathbf{x}_i \cdot \mathbf{x}_j$. Subsequently, the dot products between each $\mathbf{z}_{k=1,\ldots,M}$ can be constructed to create a new set of vectors, and the cycle repeats.

In neural networks, Eq.~\ref{eqn:dot-products} is generalized by introducing three weight matrices $\big(W_Q, W_K, W_V \big)$ with trainable parameters, and a non-linear activation function (following~\citet{vaswani_attention_2017}, we adopt the \textsc{softmax} function). The function of scaled dot product attention can be expressed compactly via matrix multiplication as
\begin{equation}
    Z = \textsc{softmax}\left(\frac{Q K^T}{\sqrt{d}}\right) V \,.
\end{equation}
Here, $Q \equiv W_Q X$, $K \equiv W_K X$, and $V \equiv W_V X$. Transformers, shown on panel (d) in Fig.~\ref{fig:arches}, are the new building blocks for designing emulators based on the dot product attention. Inside the transformer units, each output $\tilde{\mathbf{z}}_i$ contained in the matrix $Z$ is then passed to its own MLP or \textsc{ResMLP} network. The adopted Transformer implementation in this work differs from~\citet{vaswani_attention_2017} and~\citet{zhong2024attentionbased} because the MLP/\textsc{ResMLP} blocks are not identical for all output vectors $\tilde{\mathbf{z}}_i$.

 \begin{table}
     \centering
     \renewcommand{\arraystretch}{1.3}
     \begin{tabular}{lcc}
         \hline
         \textsc{HyperParameters} (HP) &  \textsc{Initial Values} 	& \textsc{Final Values} \\
         			     	    		&					& (Baseline)  \\
         \hline
         \textbf{Training} & \\
         \textsc{Batch Size} & $2500$ & $128$ \\
         \textsc{Learning Rate} & $10^{-3}$ & $10^{-3}$  \\
         \textsc{Weight Decay} ($\lambda$) & $10^{-3}$ & $10^{-3}$ \\
         \hline
         \multicolumn{3}{l}{\textbf{Architecture}: MLP and \textsc{ResMLP}}  \\
         \textsc{Width}  & $256$ & $256$ \\
         \textsc{Depth}  & $3$  & $3$\\
         \hline
         \multicolumn{3}{l}{\textbf{Architecture}: \textsc{ResBottle}} \\
         \textsc{Width} & $1024$ & $1024$ \\
         \textsc{Depth} & $3$ & $3$ \\
         \textsc{Embedding Dimension} & $256$ & $128$ \\
         \hline
         \multicolumn{3}{l}{\textbf{Architecture}: \textsc{ResTRF}} \\
         \textsc{ResMLP Width}  			& $256$ 		& $256$    \\
         \textsc{Transformer Width}  		&  $1024$  	&  $1024$ \\
         $N_\mathrm{\textsc{ResMLP}}$ 	& $3$ 		& $3$ \\
         $N_\mathrm{\textsc{Transformer}}$	& $1$ 		& $3$ \\
         $N_\mathrm{\textsc{Channels}}$ & $8$ & $32$ \\
         \hline
     \end{tabular} 
     \caption{The \textsc{Initial Values} column displays the hyperparameter values we adopt when testing how a change in one of their values affects the emulator's performance. We vary each hyperparameter individually while maintaining all others at their initial values. After we perform the hyperparameter optimization, we select the final values of hyperparameters displayed on \textsc{Final Values} column, which we adopt as the baseline configuration. The $N_\mathrm{\textsc{ResMLP}}$ and $N_\mathrm{\textsc{Transformer}}$ correspond, respectively, to the number of \textsc{ResMLP} and \textsc{Transformer} building blocks; sometimes we refer to them as the depth of the architecture.}
     \label{tab:hyperparams}
\end{table}

Besides generalizing the definition of a Transformer so each MLP block can have independent trainable parameters, this work presents the emulator displayed in Fig.~\ref{fig:full_arch} that incorporates \textsc{ResMLP} before adding transformer blocks. The proposed design contrasts with~\citet{zhong2024attentionbased}, where the self-attention-based emulator was restricted to containing only Transformer blocks, yet also draws lessons from it. As~\citet{zhong2024attentionbased} demonstrated, \textsc{ResMLP} simulates data vectors nearly at the desired level while being faster to train. As a starting point, the \textsc{ResMLP} may reduce the number of Transformer blocks needed for the emulator to achieve the expected accuracy. We implement all architectures with \textsc{PyTorch}~\citep{pytorch}.

\begin{figure}
    \centering
    \includegraphics[width=\columnwidth]{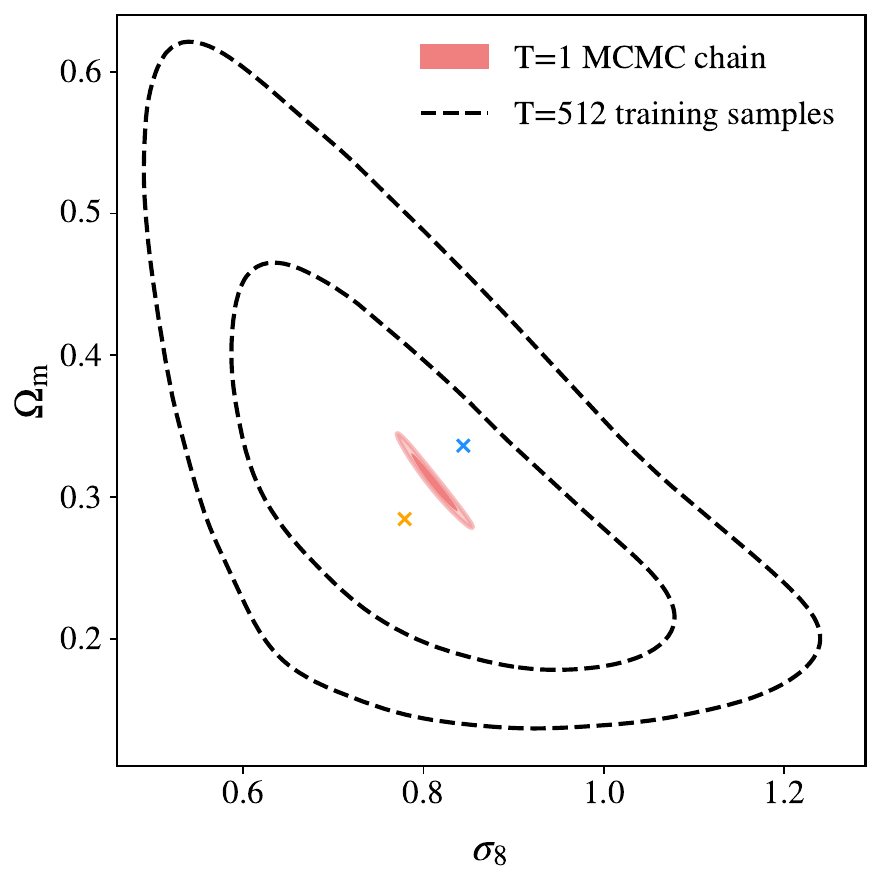}
    \caption{Comparison of the $68\%$ and $95\%$ confidence intervals between the training samples with $T=512$ and LSST Cosmic Shear forecasts without temperature rescaling (i.e., $T=1$). The crosses represent $\pm 20 \sigma$ shifts along a principal component, with $\sigma$ being the standard deviation. We label the orange cross and the blue cross as {\it cosmology 2} and {\it cosmology 3}, respectively (their values are defined in Table~\ref{tab:cosmology}). The shifts illustrate the comprehensive spread of the training data in the prior. Nonetheless, our investigation indicates that the accuracy of the \text{ResTRF} design scales well with the number of training points, $N_{\rm train}$. Therefore, the \text{ResTRF} emulator may be able to be trained on ever larger temperatures, provided that $N_{\rm train} \gtrsim 3 \times 10^6$, without further changes in its design that could include, for instance, an increase on the width of the \textsc{ResMLP} blocks.}
    \label{fig:prior}
\end{figure}

\subsection{Emulator Training}

One of the main objectives of this paper is to create emulators that can effectively cover the typical parameter range adopted by stage III/IV cosmological surveys~\citep{DES:2017qwj}. Table~\ref{tab:cosmology} displays the prior range assumed in our investigation, which followed the same $\Lambda$CDM parametrization $\boldsymbol{\theta} = \big(\log(10^{10} A_{\rm s}), n_{\rm s}, h, \omega_{\rm b} \equiv \Omega_{\rm b} h^2, \omega_{\rm c} \equiv \Omega_{\rm c} h^2 \big)$ used by \textsc{CosmoPower}~\citep{2022MNRAS.511.1771S}. With such large ranges, it is expensive to maintain the training developed in~\citet{zhong2024attentionbased}. Instead, we construct our training from the samples of a previously run LSST-Y1 MCMC chain. We then compute the covariance matrix $\mathcal{C}$ in the parameter space and create a Gaussian approximation for the likelihood distribution of the cosmological and nuisance parameters. 

The next step consists of broadening the parameter covariance by a temperature $T$, defining the new covariance by $\mathcal{C} \mapsto T\mathcal{C}$. We also widen the prior of nuisance parameters, such as shear calibration, that obey prior Gaussian distributions by setting $T_\mathrm{prior} = 100 T$. We sample from this probability distribution using the likelihood $\mathcal{L} = (\boldsymbol{\theta} - \bar{\boldsymbol{\theta}})^T (T\mathcal{C})^{-1}(\boldsymbol{\theta} - \bar{\boldsymbol{\theta}})$ with $\bar{\boldsymbol{\theta}}$ being the parameter values given in Table~\ref{tab:cosmology}. In total, we create three training sets with $N_{\rm train} = 6 \times 10^5$ (the standard set), $N_{\rm train} = 1.2 \times 10^6$ (the superior set), and $N_{\rm train} = 3 \times 10^6$ points (the enhanced set). These numbers are significantly smaller than the three training sets for cosmic shear adopted in~\citet{zhong2024attentionbased}.

We compute the data vectors using the \textsc{Cosmolike} software~\citep{2014MNRAS.440.1379E, Krause:2016jvl}, and manage the MCMC chains in this section using the \textsc{Cobaya} sampler~\citep{Torrado:2020dgo}. Both packages are integrated in the \textsc{Cobaya}-\textsc{Cosmolike} Architecture (\textsc{CoCoA})\footnote{\url{https://github.com/CosmoLike/cocoa}}. Data vector computations are trivially parallelizable with both OpenMP~\citep{dagum1998openmp} and MPI~\citep{10.5555/898758}, considerably shortening the time needed to set up the training set.

The training points displayed in Fig.~\ref{fig:prior} follow a Gaussian distribution in the parameter space with $T = 512$, chopped by the boundaries of our prior. Their extensive spread of the $T = 512$ tempered Gaussian ensures that non-Gaussian features in the initial posteriors used to create the Gaussian approximation are well sampled. It is also feasible to draw the initial samples from Fisher or DALI approximations~\citep{sellentin_breaking_2014, sellentin_non-gaussian_2015, sellentin_fast_2015}.

The coverage provided by high $T=512$ is beyond what is needed in most applications, so we also generated training sets based on Gaussian distribution with $T=128$ and $T=256$. Lower temperatures provide higher accuracy with fewer number of training models. It is worth noting that the training set in Fig.~\ref{fig:prior} has considerably broader coverage, compared to what~\citet{zhong2024attentionbased} assumed, on the cosmic shear nuisance parameters with Gaussian priors. In our previous work, the training box of such parameters was limited to $[-3.3 \, \times \, \mathrm{standard} \, \mathrm{deviation}, +3.3 \, \times \, \mathrm{standard} \, \mathrm{deviation}]$ around the central values of their Gaussian priors. 

The training data is preprocessed to enhance the training efficiency. We first preprocess the parameter vector $\boldsymbol \theta$ containing both cosmological and nuisance parameters
\begin{equation}
    \Theta_i = \frac{(\theta_i - \bar\theta_i)}{\sigma_{\theta_i}} \, .
\end{equation}
Here, $\theta_i$ is the $i$-th parameter, while $\bar\theta_i$ and $\sigma_{\theta_{i}}$ are the mean and the standard deviation of $\theta_i$. We then preprocess the cosmic shear data vectors, $\boldsymbol{y} \equiv (\xi_{+}, \xi_{-})$, as shown in Eq.~\ref{eq:dv_norm} below, normalizing them using the data vector at the fiducial cosmology, $\boldsymbol{y}_\mathrm{fid}$, and the matrix $U$ that corresponds to the change of basis matrix to the diagonal basis of the likelihood covariance $\Sigma$.
\begin{equation}\label{eq:dv_norm}
    \tilde{y}_i = \frac{(U^{-1} \boldsymbol{y})_i - (U^{-1}\boldsymbol{y}_{\mathrm{fid}})_i}{[(U\Sigma U^{-1})_{ii}]^{1/2}} \, .
\end{equation}

\begin{figure}
    \centering
    \includegraphics[width=\columnwidth]{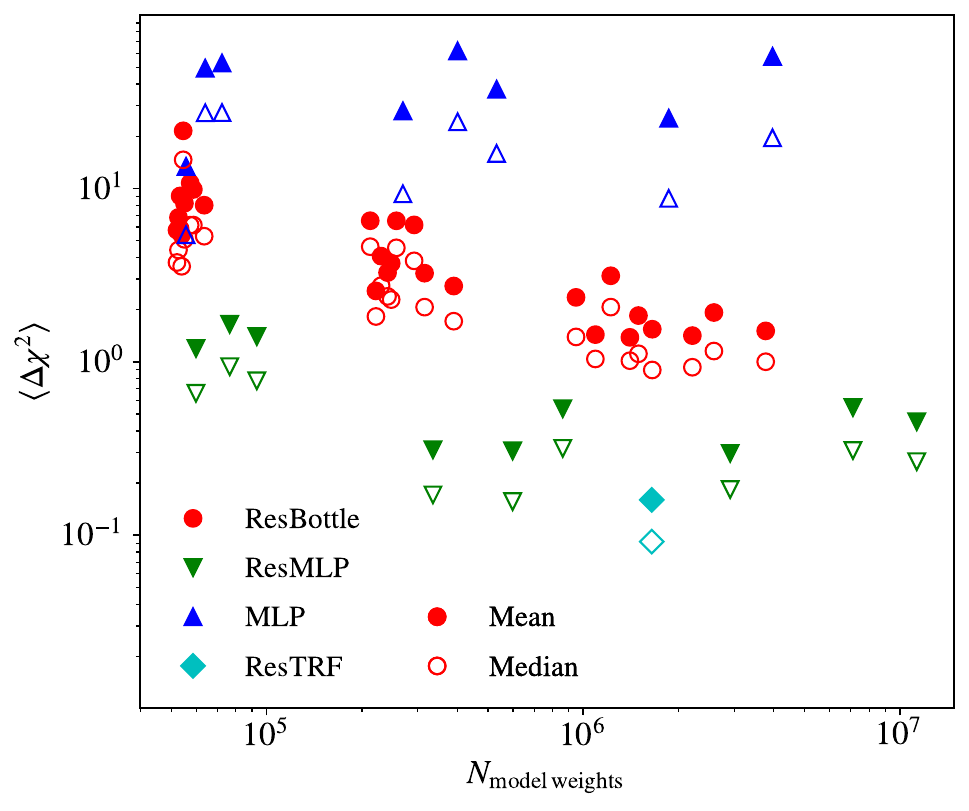}
    \caption{The mean (solid) and median (hollow) $\Delta \chi^2$ errors computed on $N_{\rm test} = 1 \times 10^4$ testing models as a function of the number of trainable parameters in the multiple neural network designs we investigated. Each neural network is trained on a $T=128$ chain and tested on a $T_{\rm test}=64$ chain. Adding more parameters to the model in all architectures does not necessarily improve the fit. Specifically, our investigation shows that the size of the training set is by far the most impactful hyperparameter on the emulator errors (see Table~\ref{tab:error_emulators}). The transformer model performs better than the others with fewer trainable parameters. We emphasize that optimizations in this plot were made for all other architectures, but not for the ResTRF. As we will see, more performance can be gained from the ResTRF architecture after hyperparameter adjustment.}
\label{fig:avg_dchi2}
\end{figure}

\begin{figure}
    \centering
    \includegraphics[width=\columnwidth]{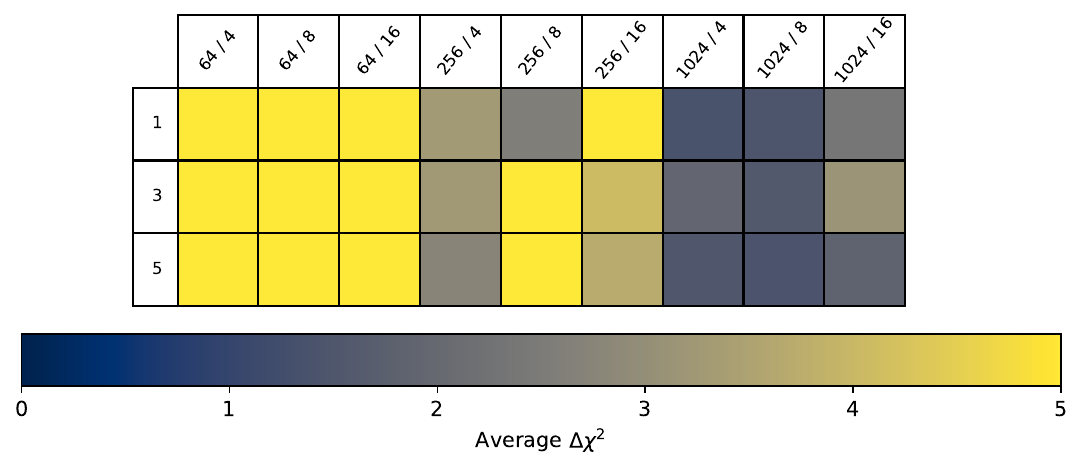}
    \caption{Detailed results on the average $\Delta \chi^2$ errors predicted by the \textsc{ResBottle} architecture trained on $T=128$ samples. The rows indicate the depth of the emulators, i.e., the number of \textsc{ResBottle} building blocks. The columns with labels $ d / N$ vary the width, $d$, of the dense layers in the \textsc{ResBottle} building blocks, and the integer $N$ that reduces the embedding dimension, $\tilde{d}$,  by $\tilde{d} = d/N$.  Across the range from $d = 64$ to $d=1024$, there is no significant reduction in the  $\langle \Delta \chi^2 \rangle$ as a function of $N$, so we can conclude that the \textsc{ResMLP} architecture does not learn unnecessary details about the $T=128$ training set. Consequently, we are not going to train and analyze the \textsc{ResBottle} design on the  $T=256$ and $T=512$ samples.}
    \label{fig:test_resbottle}
\end{figure}

\subsection{Loss Function and Optimization}

We select the $\Delta \chi^2$ between the data vector computed from the neural network emulator, $\boldsymbol{y}_{\mathrm{NN}}$, and the exact data vector computed from \textsc{CoCoA}, $\boldsymbol{y}_{\mathrm{CoCoA}}$, to be the loss function, $L(\boldsymbol{w})$, of our model. 
\begin{align}
\Delta \chi^2 & \equiv  \big(\boldsymbol{y}_{\mathrm{NN}} - \boldsymbol{y}_{\mathrm{CoCoA}} \big)^T \boldsymbol\Sigma^{-1} \big(\boldsymbol{y}_{\mathrm{NN}} - \boldsymbol{y}_{\mathrm{CoCoA}} \big) \\
    L(\boldsymbol{w}) &= \langle \Delta \chi^2  \rangle \,.
\end{align}
Here, $\boldsymbol{w}$ is a vector containing all trainable parameters, $\langle \,\rangle$ denotes the sample mean, and $\boldsymbol\Sigma$ is the cosmic shear covariance matrix computed with \textsc{CosmoCov}~\citep{Krause:2016jvl,Fang2020}. This covariance accounted for the Gaussian, connected non-Gaussian, and super-sample effects.

Similar to~\citet{zhong2024attentionbased}, the sample mean does not have any weight that would give preference to samples closer to the fiducial cosmology, nor any procedure to remove outliers. While training, we use the $L^2$-norm of the model weights to regulate the network and prevent overfitting~\citep{10.1145/1015330.1015332}. In \textsc{PyTorch}, this is done via a weight decay parameter that acts as a multiplicative constant on the $L^2$-norm.

At the beginning of training, a high learning rate allows the network to explore a large volume of the weight space~\citep{bengio2012practical}. However, it can also prevent the network from descending into the minima of the loss function. On the other hand, a low learning rate can prevent the network from finding the global minimum of the loss function. The adaptive learning rate (ALR) will decrease the learning rate over the course of the training, getting the benefit of weight-space exploration without preventing descent into minima. As such, we adopted the \textsc{Adam} optimizer with adaptive learning rate (ALR) to train the network. The ALR was implemented via \textsc{PyTorch}'s \textit{reduce LR on plateau} scheduler, which decreases the learning rate by a factor of ten when the validation loss plateaus for ten epochs. We set the minimum learning rate to be $10^{-8}$.

\subsection{Choice of Hyperparameters}

Hyperparameters can greatly affect the training and generalization of a neural network. The relevant hyperparameters in our models are the model width, depth (number of blocks), learning rate, weight decay, batch size, number of channels in the attention block, and the number of training samples. This section explores the variation of each of these parameters while maintaining all others at the initial values described in Table~\ref{tab:hyperparams}. Unless otherwise specified, the quoted numbers were calculated comparing $\textsc{CoCoA}$ with the \textsc{ResTRF} emulator. We trained the emulators, except for $\textsc{ResBottle}$, assuming $T=512$, and created an independent set of $N_{\rm test} = 1 \times 10^4$ testing samples that assumed $T_{\rm test}=256$. Setting the validation temperature to be half the value used in training reduces spurious effects that would inevitably arise from the degradation of emulators near the edge of the training samples. Table~\ref{tab:hyperparams} also the defined baseline set that summarizes our findings.

\textbf{Activation function}: Neural networks can learn non-linear mappings by the inclusion of activation functions. These non-linear functions generally fall into, but are not limited by, two categories: the \textit{rectified linear} functions and the \textit{sigmoid} functions. We test the Rectified Linear Unit (\textsc{ReLU}),
\begin{equation}\label{eq:relu}
    \mathrm{\textsc{ReLU}}(x) = \left\{ \begin{array}{lcl}
        x & , & x>0 \\
        0 & , & x\leq0
    \end{array} \right. \,,
\end{equation}
and the hyperbolic tangent (\textsc{Tanh}). To avoid saturating the \textsc{Tanh} activation function, we employ an affine normalization of the form $gx+b$, with $g$ $b$ being numbers, between each layer before activation. We chose \textsc{Tanh} because it is an antisymmetric function, adding another distinction between it and \textsc{ReLU}. Since components of the preprocessed data vector can be negative, the model may behave better with non-zero activation functions for negative values. Using the \textsc{ReLU} activation function outperforms \textsc{Tanh} by a significant margin in the standard training set. The mean (median) $\langle \Delta \chi^2\rangle$ for \textsc{ReLU} and \textsc{Tanh} were $\langle \Delta \chi^2\rangle({\rm AF}=\textsc{ReLU}) = 3.4\,(1.3)$ and $\langle \Delta \chi^2 \rangle({\rm AF}=\textsc{Tanh}) = 6.4\,(2.7)$, respectively. However, as we will see, the \textsc{Tanh} scales better with the size of the training set, and the conclusion flips in the enhanced set.

\textbf{Weight decay}: The weight decay $\lambda$ penalizes the model for highly relying on a certain network region. This unwanted reliance is determined by the weights squared ($L^2$ norm) value, as large weights indicate a high dependence on some particular neuron. We use a baseline of $\lambda=10^{-3}$ and test values of $\lambda=10^{-1}$ and $\lambda=10^{-5}$.  Generally, we find that a lower weight decay of $\lambda=10^{-5}$ gives the lowest mean and highest median at $\langle \Delta \chi^2\rangle({\lambda=10^{-5}}) = 14.1$  and $\langle \Delta \chi^2\rangle_{\rm median}(\lambda=10^{-5}) = 7.0$, respectively. The $\lambda=10^{-3}$ and $\lambda=10^{-1}$ have mean (median) $\langle \Delta \chi^2\rangle(\lambda=10^{-3}) = 6.4 \, (2.7)$ and $\langle \Delta \chi^2\rangle(\lambda=10^{-1}) = 8.4\,  (2.9)$. Even though the median $\Delta\chi^2$ is similar in both cases, we adopted $\lambda=10^{-3}$ because it gives the lowest mean $\Delta \chi^2$ indicating it has less high-$\Delta\chi^2$ outliers.

\textbf{Architectures}: We examine how varying the building blocks' width and depth affected the validation loss. We set their widths ranging from $64$ to $1024$ and their depths ranging from $1$ to $5$. On the \textsc{ResBottle} design, we study the choices $\tilde{d} = d/2$, $\tilde{d} = d/4$, and $\tilde{d} = d/8$ for the embedding dimension $\tilde{d}$. The comparison between \textsc{ResBottle} and \textsc{ResMLP} inspects whether the \textsc{ResMLP} learns unnecessary information about the training set. Figs.~\ref{fig:avg_dchi2} and~\ref{fig:test_resbottle} show that a bottlenecked layer degrades the emulator's accuracy in all tested configurations when trained on $T=128$ samples. Given that higher temperatures are more challenging to emulate, we conclude that the \textsc{ResMLP} architecture does not learn unnecessary details about training sets with $T \geq 128$.

 \begin{table}
     \centering
     \renewcommand{\arraystretch}{1.3}
     \begin{tabular}{|l|cc|}\hline
           & $N_\mathrm{\textsc{TRF}}=1$ & $N_\mathrm{\textsc{TRF}}=3$ \\\hline
         $N_\mathrm{\textsc{ResMLP}}=0$ & $12.6\,(5.9)$ & $7.5\,(5.2)$   \\
         $N_\mathrm{\textsc{ResMLP}}=1$ & $7.4 \,(3.5)$ & $4.7\,(2.6)$  \\
         $N_\mathrm{\textsc{ResMLP}}=3$ & $6.4 \,(2.7)$ & $4.9\,(2.3)$   \\
         $N_\mathrm{\textsc{ResMLP}}=4$ & $8.1 \,(4.7)$ & $6.0\,(3.1)$   \\\hline
     \end{tabular} 
     \caption{The mean (median) $\Delta \chi^2$, which represents the \textsc{ResTRF} discrepancies against \textsc{CoCoA}, as a function of the number $N_\mathrm{\textsc{ResMLP}}$ of \textsc{ResMLP} blocks and in the number $N_\mathrm{\textsc{TRF}}$ of transformer blocks in the design described in Fig.~\ref{fig:full_arch}. We keep all remaining hyperparameters to the \textsc{initial values} shown in Table~\ref{tab:hyperparams}, trained the emulator assuming $T=512$, and compute the averages from $N_{\rm test} = 1 \times 10^4$ test points drawn from the $T_{\rm test} = 256$ distribution. We highlight the similarities in the results from $N_\mathrm{\textsc{Transformer}}=3$ and $N_\mathrm{\textsc{ResMLP}}=1,3$. While the mean is lower with one \textsc{ResMLP}, the median is larger, which can be due to the impact of outliers on training.
}
\label{tab:architecture_variations}
\end{table}

 \begin{table}
     \centering
     \renewcommand{\arraystretch}{1.3}
     \begin{tabular}{|l|cc|}\hline
            & \textsc{Learning rate}=$10^{-3}$ & \textsc{Learning rate}=$10^{-4}$ \\\hline
         \textsc{Batch Size}=$128$  & $2.7\,(1.2)$ & $5.3\,(2.1)$ \\
         \textsc{Batch Size}=$256$  & $2.4\,(1.1)$ & $2.8\,(1.1)$ \\
         \textsc{Batch Size}=$1024$ & $18.6\,(6.1)$ & $9.2\,(4.4)$ \\
         \textsc{Batch Size}=$2500$ & $6.3\,(2.7)$ & $12.6\,(5.9)$ \\\hline
     \end{tabular}
     \caption{The mean (median) $\Delta \chi^2$, which represents the \textsc{ResTRF} discrepancies against $\textsc{CoCoA}$, as a function of the batch size and learning rate. We kept all remaining hyperparameters to the \textsc{initial values} shown in Table~\ref{tab:hyperparams}, trained the emulator assuming $T=512$, and computed the averages from $N_{\rm test} = 1 \times 10^4$ test points drawn from the $T_{\rm val} = 256$ distribution.
}
\label{tab:lr_bs}
\end{table}

 \begin{table}
     \centering
     \renewcommand{\arraystretch}{1.3}
     \begin{tabular}{|l|ccc|}\hline
            & $N_\textsc{CH}$=8 & $N_\textsc{CH}$=32 & $N_\textsc{CH}$=128 \\\hline
         \textsc{TRF Width}=64   & $11.9\,(5.7)$ & $16.7\,(7.8)$ & $-$\\
         \textsc{TRF Width}=256  & $5.9\,(2.5)$ & $25.0\,(11.5)$ & $14.0\,(6.2)$ \\
         \textsc{TRF Width}=1024 & $6.4\,(2.7)$ & $2.7\,(1.0)$ & $4.8\,(1.7)$ \\
         \textsc{TRF Width}=2048 & $21.8\,(6.7)$ & $4.6\,(2.0)$ & $2.8\,(1.0)$ \\\hline
     \end{tabular}
     \caption{The mean (median) $\Delta \chi^2$, which represents the \textsc{ResTRF} discrepancies against $\textsc{CoCoA}$, as a function of the number of channels, $N_\textsc{CH}$, and the size of input of the transformer block (TRF). We keep all remaining hyperparameters to the \textsc{initial values} shown in Table~\ref{tab:hyperparams}, train the emulator assuming $T=512$, and compute the averages from $10,000$ test points drawn from the $T_{\rm test} = 256$ distribution.}
     \label{tab:trf_width_channels}
\end{table}

\begin{figure*}
	\centering
	\includegraphics[width=\textwidth]{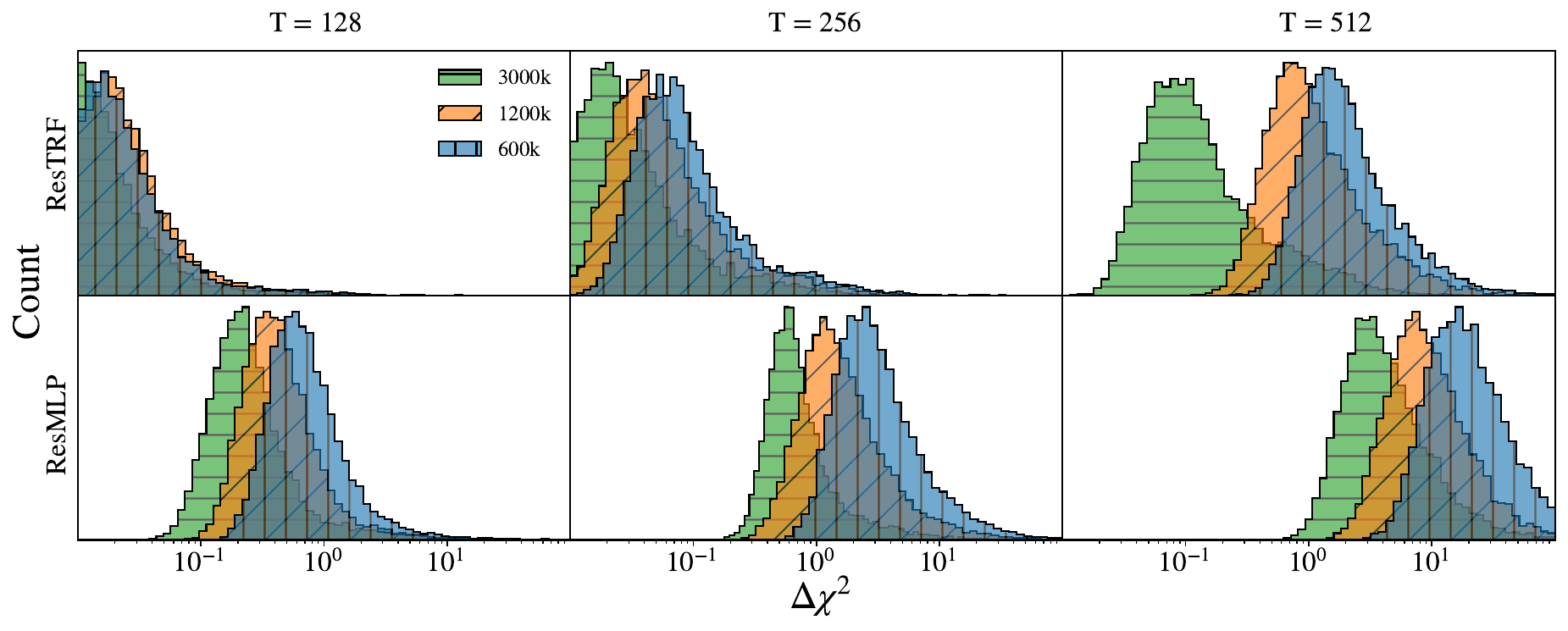}
	\caption{The distribution of cosmic shear $\Delta \chi^2$ differences between \textsc{ResTRF} and \textsc{ResMLP} emulators (\textit{top} and \textit{bottom} panels) against the exact $\textsc{CoCoA}$ data vectors. The training temperatures increase from $T=128$ (\textit{left-panels}) to $T=256$ (\textit{mid-panels}) and $T=512$ (\textit{right-panels}). The green histogram with horizontal lines corresponds to our results when setting the number of training models at $N_{\rm train} = 6 \times 10^5$ (standard set). Additionally, the orange with diagonal lines and blue with vertical lines histograms assume $N_{\rm train} = 1.2 \times 10^6$ (superior set) and $N_{\rm train} = 3 \times 10^6$ (enhanced set), respectively. For all scenarios, we computed the $\Delta \chi^2$ distribution from $N_{\rm test} = 1 \times 10^4$ testing points distributed according to a normal likelihood distribution on the cosmological parameters with a covariance that was tempered by the reduced testing temperature $T_{\rm test} = T/2$. We also imposed the hard priors shown in Table~\ref{tab:cosmology} in the tempered MCMC that distributed the testing points. For both \textsc{ResTRF} and \textsc{ResMLP} emulators, the size of the training set plays a considerable role in their overall accuracy, even when considering the massive number of models in the baseline training. The trending displayed in these panels suggests that $N_{\rm train} \sim 10^7$ could reduce the \textsc{ResMLP} fraction of validation points \texthtbardotlessj($\Delta \chi^2 > 1.0$) to single digits at $T=512$. However, the better-suited attention-based \textsc{ResTRF} emulator could push these same models toward having \texthtbardotlessj$(\Delta \chi^2 > 0.2) < 0.1$ with a similar training size. Finally, the \textsc{ResTRF} emulator predicted a narrow mean/median $\Delta \chi^2$ gap between the standard and enhanced training on $T=128$ and $T=256$. This suggests that further improvements are hard to achieve when the emulator accuracy reaches mean/median $\langle \Delta \chi^2\rangle \approx 0.1$. In this case, a more effective usage of such large sets would involve training the emulator at even higher temperatures (e.g., $T=512$).
	\label{fig:num_training}
}
\end{figure*}

The comparison between architectures is summarized in Fig.~\ref{fig:avg_dchi2}, which displays the mean and median $\Delta \chi^2$ as a function of the number of trainable parameters. Fig.~\ref{fig:avg_dchi2} demonstrates that the transformers-based model provides considerable improvement over all other designs, even when they have more trainable parameters. In the end, we set a fixed input size of $256$ and $1024$ on the \textsc{ResMLP} and transformer blocks in the \textsc{ResTRF} emulator, respectively.

Regarding the number of $\textsc{ResMLP}$ ($N_\mathrm{\textsc{ResMLP}}$) and transformer ($N_\mathrm{\textsc{TRF}}$) blocks. We test combinations of $N_\mathrm{\textsc{ResMLP}}  = 0,1,3,4$ \textsc{ResMLP} blocks and $N_\mathrm{\textsc{TRF}} = 1,3$ transformers blocks. We find consistent improvement when having $3$ transformer blocks compared to $1$. In contrast, there is no consistent improvement when adding more \textsc{ResMLP} blocks before the transformer. The peak performance is at $N_\textsc{ResMLP}=3$, while $N_\textsc{ResMLP}=1$ can get similar performance. This might be caused by an overparameterization of the model, as each \textsc{ResMLP} provides an additional $1.3 \times 10^5$ parameters. In contrast, each transformer adds only $4 \times 10^4$ parameters when using the baseline hyperparameters.

When keeping the number of transformers at $N_\mathrm{\textsc{TRF}} = 3$, we find the median $\Delta \chi^2$ is the lowest at $\langle \Delta \chi^2 \rangle_{\rm median} = 2.3$ with $N_\mathrm{\textsc{ResMLP}} = 3$ blocks. The worst agreement happened when we adopt $N_\mathrm{\textsc{ResMLP}}= 0$, resulting in $\langle \Delta \chi^2 \rangle_{\rm median} = 5.9$ with one transformer; this case also shows a large mean pointing towards an excess of outliers. Compared to $N_\mathrm{\textsc{ResMLP}}= 3$, the median $\Delta\chi^2$ increased by $\sim$ 1-2 with $N_\mathrm{\textsc{ResMLP}} = 4$. The behavior when using $N_\mathrm{\textsc{TRF}} = 1$ was that $\langle \Delta \chi^2 \rangle_{\rm median} \approx 2.7$ with three residual blocks, and the agreement is more dependent on the number of residual blocks compared with $N_\mathrm{\textsc{TRF}} = 3$. These results are summarized in Table~\ref{tab:architecture_variations}.

 \begin{table}
     \centering
     \renewcommand{\arraystretch}{1.05}
     \begin{tabular}{|l|c|c|c|c|}\hline
         & $N_\mathrm{train} / 10^3$ &\texthtbardotlessj$({\Delta \chi^2 > 1})$ & \texthtbardotlessj$(\Delta \chi^2 > 0.2)$ \\\hline
        \textsc{ResMLP128}   & 600  & 0.26     & 0.99    	\\
        \textsc{ResTRF128}  & 600  & <0.01     & 0.03      	\\ \hline 
        \textsc{ResMLP256}  & 600  & 0.97     & \cellcolor{red!25} 1  		\\
        \textsc{ResTRF256}  & 600  & 0.03    & \cellcolor{blue!25} 0.17    	\\ \hline 
        \textsc{ResTRF512}  & 600  & 0.81     &  	1	\\ \hline \hline 
        \textsc{ResMLP128}  & 1200  & 0.10   & 0.90    	\\
        \textsc{ResTRF128}  & 1200  & <0.01 & 0.03       	\\ \hline 
        \textsc{ResMLP256}  & 1200  & 0.71   & \cellcolor{red!25} 1  		\\
        \textsc{ResTRF256}  & 1200  & 0.02   & \cellcolor{blue!25} 0.12   	\\ \hline 
        \textsc{ResTRF512}  & 1200  & 0.48   & 0.99   	\\ \hline \hline 
        \textsc{ResMLP128}  & 3000  & 0.07   & 0.57    	\\
        \textsc{ResTRF128}  & 3000  & <0.01 & 0.02    	\\ \hline 
        \textsc{ResMLP256}  & 3000  & 0.22   & \cellcolor{red!25} 1  		\\
        \textsc{ResTRF256}  & 3000  & \cellcolor{yellow!25} <0.01   & \cellcolor{blue!25} 0.06     	\\ \hline
        \textsc{ResMLP512}  & 3000  & 0.99   &  1  		\\
        \textsc{ResTRF512}  & 3000  &  0.06  & 0.29  	\\ \hline
    \end{tabular}
	\caption{Fraction of validation points (\texthtbardotlessj) with errors larger than the thresholds $\Delta \chi^2 = \{0.2, 1\}$. According to the Dark Energy Survey, $\Delta \chi^2 > 1$ errors require mitigation via importance sampling, while $\Delta \chi^2 < 0.2$ are considered insignificant~\citep{DES:2021rex}. The postfix number in the emulator name indicates the training temperature, and $N_\mathrm{train}$ is the number of training points. In all cases, the validation temperature is reduced by half to avoid the quoted numbers being dominated by outliers near the training boundaries. The red background highlights the inability of the \textsc{ResMLP} emulator to increase the number of validation points meeting the accuracy threshold $\Delta \chi^2 = 0.2$ on $T=256$, even when quintupling the number of training points. Conversely, the \textsc{ResTRF} design reduces \texthtbardotlessj$(\Delta \chi^2 > 0.2)$ from the already impressive $0.17$ to $0.06$, when $N_\mathrm{train}$ was increased from $N_\mathrm{train} = 6 \times 10^5$ to $N_\mathrm{train} = 3 \times 10^6$ (blue background). Even on $T=512$, the  \textsc{ResTRF} reduced the fraction of points above $\Delta \chi^2 = 0.2$ to $0.29$ when trained with three million points. Finally, the yellow background highlights the remarkable accuracy of the  \textsc{ResTRF} emulator trained with 3 million points at $T=256$, with only $1\%$ of the validation models having $(\Delta \chi^2 > 1)$.}
	\label{tab:error_emulators}
\end{table}

\textbf{Batch Size and Learning Rate}: The batch size and learning rate are considered together, as both hyperparameters affect the minimization process in the \textsc{Adam} optimizer~\citep{smith_dont_2018,hoffer_train_2018}. \textsc{Adam} updates the parameters by averaging the gradient over batches~\citep{kingma_adam_2017}. As such, the batch size affects the stability of the direction of the step. On the other hand, the learning rate acts as a step size. Thus, both affect the ability of the optimizer to converge to the global minimum. We consider batch sizes of $2500$, $1024$, $256$, and $128$, and initial learning rates of $10^{-3}$ and $10^{-4}$. 

In both cases of starting learning rate of $10^{-3}/10^{-4}$, the mean and median $\Delta \chi^2$ achieved by the emulator generally reduce when the batch size is reduced, except when going from 256 to 128 (there was also a bump from 2500 to 1024 on learning $10^{-3}$). This might indicate that small batch sizes make the emulator more generalizable. Since we use a learning rate scheduler to decrease the learning rate when the validation loss plateaus, starting with a larger learning rate is typically better. Therefore, we set a batch size of $256$ and a starting learning rate of $10^{-3}$. Table~\ref{tab:lr_bs} summarizes our findings.

\section{Emulator Validation}

The comparison against \textsc{CoCoA} is accomplished at three levels, the first one being the computation of the $\Delta \chi^2$ distribution on a testing set comprising $N_{\rm test} = 10^4$ points distributed following a normal likelihood distribution on the cosmological and nuisance parameters with a tempered covariance by the reduced temperature $T_{\rm test}= T/2$. We also impose the hard priors shown in Table~\ref{tab:cosmology} in the tempered MCMC that distributed the testing points. The downsized testing temperature prevents regions in parameter space with sparse training coverage from biasing our testing statistics.  

In level one, we also analyze the fraction of testing points with $\Delta \chi^2$ errors above the two representative thresholds $\Delta \chi^2 = (0.2, 1)$. According to the Dark Energy Survey year three analysis, $\Delta \chi^2 > 1$ errors require mitigation via importance sampling, while $\Delta \chi^2 < 0.2$ are considered insignificant~\citep{DES:2021rex}. Additionally,~\citet{Campos:2022gtg} has shown on simulated Dark Energy Survey year three cosmic shear data that discrepancies between intrinsic alignment models at the order of $\Delta \chi^2 \sim 0.2$ are significantly less likely, by an approximate factor 10, to induce parameter biases in the $S_8$-$\Omega_{\rm m}$ plane exceeding $0.3$ times their standard deviation compared to $\Delta \chi^2 \sim 1$. This finding corroborates our push towards an overall $\Delta \chi^2 \lesssim 0.2$ accuracy in all attention-based emulators.

Levels two and three are practical tests that demonstrate the ability of our emulator to reproduce posteriors and Bayesian evidence. In both cases, we shift the cosmology at which the posteriors are centered, testing our emulator's ability to cover large volumes of parameter space. The shifts are done along the most constraining direction of the posterior in the $\sigma_8$-$\Omega_m$ plane. We compare our results to the posteriors and evidences obtained using \textsc{CoCoA}.

\begin{figure*}
        \centering
        \includegraphics[width=\textwidth]{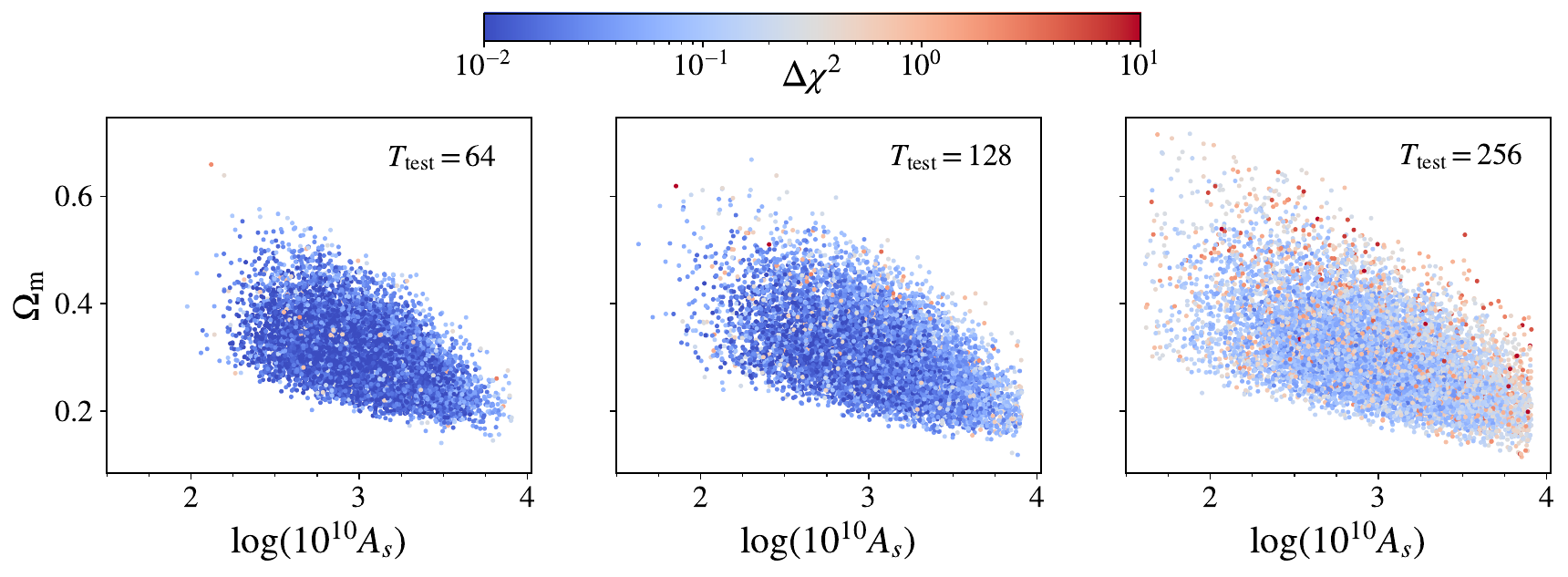}
        \caption{The projected spatial distribution of $\Delta \chi^2$ emulator errors, computed on $N_{\rm test} = 1 \times 10^4$ testing points distributed according to a normal likelihood distribution on the cosmological parameters with a covariance that was tempered by the reduced testing temperature $T_{\rm test} = T/2$. We also imposed the hard priors shown in Table~\ref{tab:cosmology} in the tempered MCMC that distributed the testing points. This reduced testing temperature helps prevent regions in parameter space with sparse training coverage from biasing our testing statistics. The left, middle, and right panels were trained on the standard $N_{\rm train} = 6 \times 10^5$, superior $N_{\rm train} = 1.2 \times 10^5$, and enhanced $N_{\rm train} = 3 \times 10 ^6$ sets, respectively. In all cases, we set the hyperparameters to the baseline configuration displayed in Table~\ref{tab:hyperparams}. In many directions in parameter space, the prior and validation set hit the prior boundaries, which degrades the emulator fit, as seen on the right panel. Specifically, the mean (median) $\Delta \chi^2$ decreases from $\langle \Delta \chi^2 \rangle = 0.33 \, (0.11)$ to $\langle \Delta \chi^2 \rangle = 0.24 \, (0.08)$, at  $T_{\rm test}=256$, when we reduce the prior boundaries by $10\%$ in each direction. In the future, we may expand the training hard prior boundaries by a few percent in each parameter direction to prevent this problem.  For most applications, particularly those involving LSST-Y1 forecasts, the $T_{\rm test} = 64$ provides sufficient coverage and excellent accuracy.}
        \label{fig:val_diff_temperatures}
\end{figure*}

\subsection{First Validation Level: $\Delta \chi^2$ Errors}

Fig.~\ref{fig:num_training} shows that the \textsc{ResTRF} is several times more accurate than \textsc{ResMLP} across all temperatures and training sizes. Both on $T= 128$ and $T= 256$, the standard training is sufficient to shift the median to $\langle \Delta \chi^2 \rangle_{\rm median} \sim 0.1$. Specifically, just $17\%$ of the validation points exhibit  $\Delta \chi^2 > 0.2$ on  $T=256$. Increasing the training sample to the superior and enhanced sets reduced this fraction to $12\%$ and $6\%$, respectively, a modest improvement indicating that the emulator learning plateaus once it successfully reduced its median errors below $0.1$. Only the $T=512$ offers a real challenge to \textsc{ResTRF} trained on the standard set; even in this case, the median error is still around one.

In all temperatures, the accuracy of the \textsc{ResTRF} with a \textsc{ReLU} activation does not significantly improve when we increase the number of training points from the standard to the superior set. However, Fig.~\ref{fig:num_training} clearly shows that the same is not valid with the \textsc{Tanh} activation, where there is a consistent and considerable improvement as the number of training samples increases. The \textsc{ResTRF} and \textsc{ResMLP} predict similar gaps between the standard and superior training sets. When training \textsc{ResTRF} with the enhanced set, the proportion of testing points with $\Delta \chi^2 > 0.2$ drops from $99\%$ to only $29\%$ on $T=512$; there is also an order of magnitude improvement in the median $\Delta \chi^2$ from the standard to the enhanced set on $T=512$. Overall, the \textsc{ResMLP} shows a more predictable scaling, but \textsc{ResTRF} demonstrates that it can still be significantly improved, decreasing its errors from $\Delta\chi^2 \sim 1$ to $\Delta\chi^2 \sim 0.1$, when a few million models are used in training.

The $\Delta \chi^2$ distribution of the \textsc{ResTRF} on $T=512$ resembles the \textsc{ResMLP} predictions on $T=128$. This similarity exemplifies our experience that introducing transformer blocks allowed temperatures to be increased by a factor of a few without accuracy degradation. On a fixed temperature, the standard \textsc{ResTRF} consistently outperforms the enhanced \textsc{ResMLP}. Nevertheless, the additional models on the enhanced training set allow the transformer-based emulator to predict virtually every testing point with $\Delta \chi^2 < 1$ accuracy on $T=256$. Indeed, the enhanced \textsc{ResTRF} is close to universally guaranteeing the much more stringent $\Delta \chi^2 < 0.2$ threshold. On the other hand, the enhanced \textsc{ResMLP} predicted $22\%$ of the testing points with errors larger than $\Delta \chi^2 > 1$.

From Table~\ref{fig:num_training}, we see that \textsc{ResMLP} cannot emulate $90\%$ of the models at the $\Delta \chi^2 \sim 0.2$ accuracy level for temperatures $T \gg 128$. This conclusion should hold even when considering training sizes at the order of $N_{\rm train} \lesssim 10^7$. Specifically, the \textsc{ResMLP256} predicted nearly all testing points with $\Delta \chi^2 > 0.2$ when trained on the enhanced set, even though only $22\%$ of them had $\Delta \chi^2 > 1$. Based on practical considerations about memory, CPU, and GPU consumption, we set ten million as the maximum reasonable number of training points. However, one may push training to tens of millions of datavectors, and then, finally, the \textsc{ResMLP} might reach the $\Delta \chi^2 \lesssim 0.2$ accuracy goal. Our comparison tests how fast these designs can learn to simulate the cosmic shear data vector; we did not study how the architectures perform on (almost unlimited) $N_{\rm train} \gg 10^6$ training sizes.

We define the limiting emulator temperature, $T_{\rm max}$, as the maximum temperature at which the emulator can reduce the fraction of testing points with $\Delta \chi^2 > 0.2$ to below $10\%$, assuming training sizes at the order of $N_{\rm train} \lesssim 10^7$. Our analysis suggests that the limiting temperature for the \textsc{ResMLP} and \textsc{ResTRF} architectures are 128 and at least 512, respectively. This factor of four, possibly eight, matters, as emulators will be most helpful in inferences involving additional two-point functions and parameters that model new physics in the dark sector and systematics. The potentially much larger parameter space may reduce $T_{\rm max}$ from our quoted values. In the limit $T_{\rm max} \ll 64$, retraining may become frequently necessary, and in this case, the training method described in~\citet{Boruah:2022uac} becomes advantageous. 

For example, inferences that involve the ten two-point correlation functions that can be generated by cross-correlating galaxy shapes, galaxy positions, CMB lensing, and the thermal Sunyaev-Zel’dolvich introduce dozens of nuisance parameters~\citep{Fang:2023efj}. Higher order galaxy biases, modeled using Hybrid Effective Field theories, add three nuisance parameters per lens tomographic redshift bins~\citep{Modi:2019qbt, Kokron:2021xgh, Nicola:2023hsd}. The recent~\citet{DES:2021zxv} analysis exemplifies the extensive number of nuisance parameters required to model systematics in studies that include small scales. Modeling these correlation functions in the context of the more precise LSST year ten will certainly require a lower $T_{\rm max}$. 

However, there are limitations to our training approach. Relying on the mean $\Delta\chi^2$ as the loss function means the training is prone to outliers. These points disproportionately contribute to the mean, forcing the emulator to learn them at the expense of the remaining points. We examine this effect by removing five points from each batch that contributed the most to the loss. In the standard training set with the baseline transformer hyperparameters at a temperature of $T=512$, the outlier removal results in $\langle \Delta\chi^2 \rangle_{\rm median} = 0.03$ with only \texthtbardotlessj$(\Delta \chi^2 > 0.2) = 0.12$. In this case, the mean $\Delta\chi^2$ is no longer meaningful to quantify the agreement between the emulator and \textsc{CoCoA}. A detailed analysis of outlier mitigation will be published in part III. Despite this, the \textsc{ResTRF} appears significantly more resilient to outliers than the other architectures.

\subsection{Second Validation Level: Parameter Shifts}

The second validation level involves direct comparisons between \textsc{CoCoA} and the emulators at the posterior level. All posterior distributions in this section are computed via MCMC simulations, with \textsc{CoCoA} employing an Adaptive Metropolis Hasting (AMH) algorithm~\citep{Lewis:2013hha, Torrado:2020dgo}. In the AMH sampler, we terminate the chains once the Gelman-Rubin convergence diagnostic reached $R-1<0.01$ for the means and $R-1<0.12$ for the standard deviations.

\begin{figure}
    \centering
    \includegraphics[width=\columnwidth]{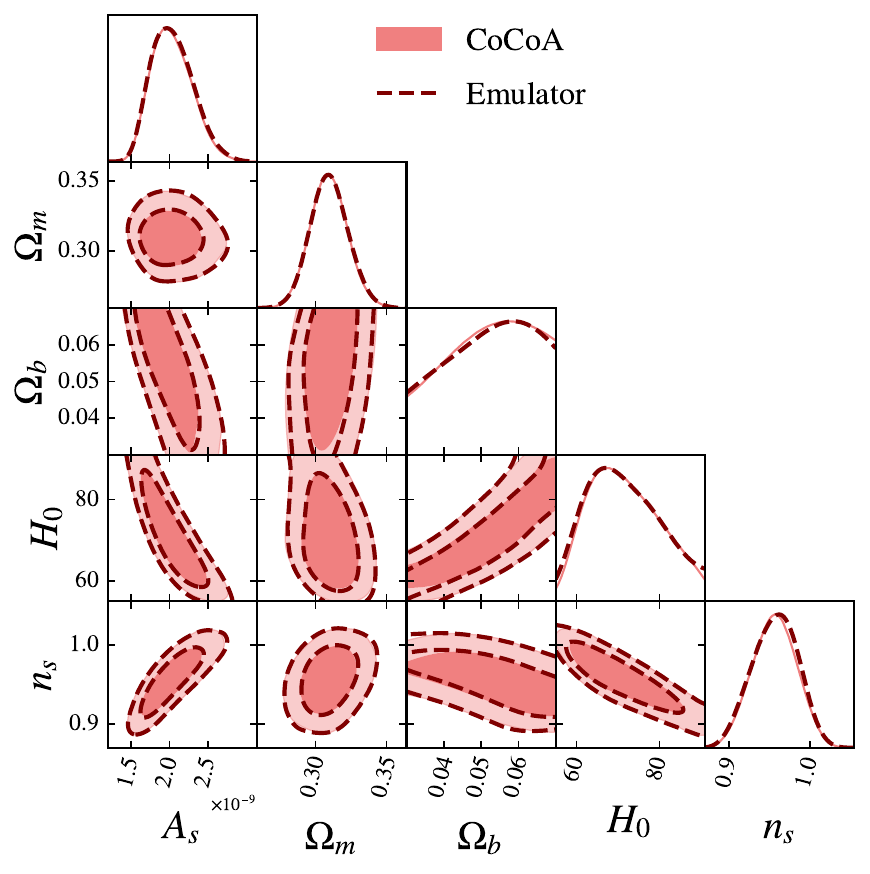}
    \caption{Comparison at the level of the posterior between the \textsc{ResTRF256} emulator, trained on the enhanced set ($N_\mathrm{train} = 3 \times 10^6$), against \textsc{CoCoA}. The emulator chain is generated using \textsc{Cobaya} with a convergence criterion of $R-1=0.01$ of the means. In this basic test, we highlight the marginal posteriors of all cosmological parameters, including the ones that are weakly constrained. Even though we do not observe boundary issues, we plan to expand the hard prior adopted on the emulator training in future work. The cosmology that created the fiducial synthetic cosmic shear data vector is shown in Table~\ref{tab:cosmology}. As expected, the small $\Delta \chi^2 \ll 1$ errors of the \textsc{ResTRF256} emulator convert into an excellent match at the posterior level.}
    \label{fig:posteriors}
\end{figure}

\begin{figure}
    \centering
    \includegraphics[width=0.895\columnwidth]{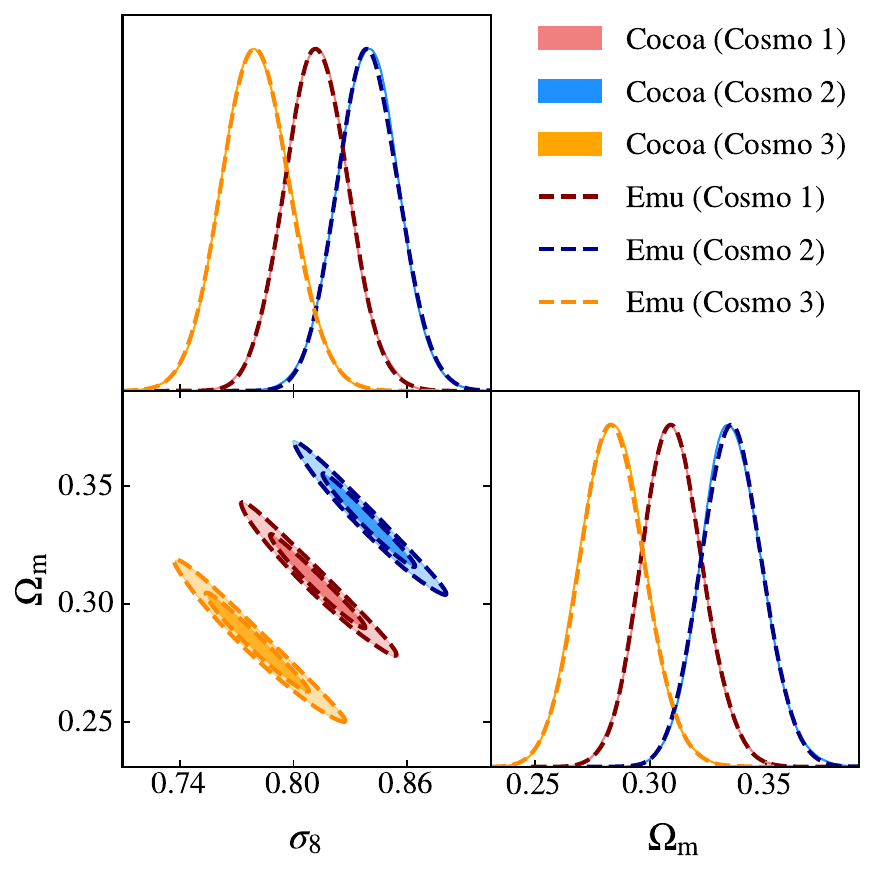}
    \includegraphics[width=0.895\columnwidth]{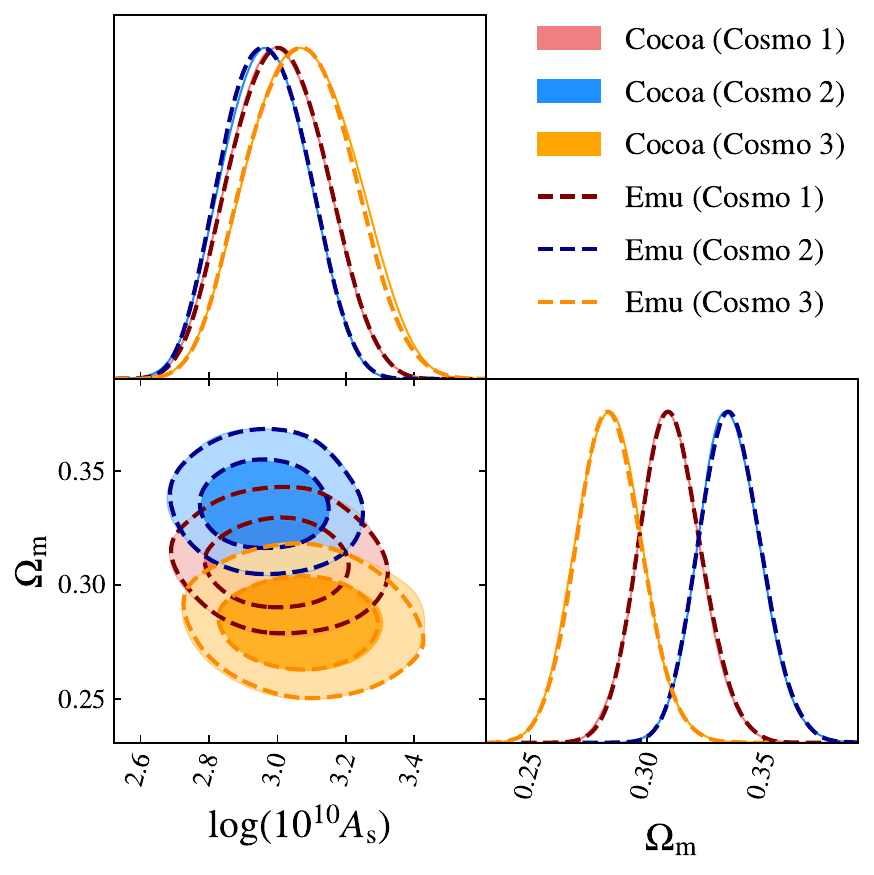}
    \caption{Comparison at the level of the posterior between the \textsc{ResTRF512} emulator, trained on the enhanced set ($N_\mathrm{train} = 3 \times 10^6$), against the exact \textsc{CoCoA} calculation on three different cosmologies. The first cosmology, labeled \textsc{Cosmo 1}, corresponds to the fiducial model given in Table~\ref{tab:cosmology}, while the cosmologies $2$ and $3$ were generated by shifting the parameters along the first principal component (PC) in the $\sigma_8$-$\Omega_{\rm m}$ plane by $\pm 20 \sigma_\mathrm{PC}$ (the exact parameter values are shown in Table~\ref{tab:cosmology})  ({\it top panel}). Our emulator does not accept $\sigma_8$ as an input parameter, so we show the same chains in the $\log(10^{10} A_{\rm s})$-$\Omega_{\rm m}$ plane ({\it bottom panel}). Although these PC shifts map into more moderate changes in the $\log(10^{10} A_{\rm s})$-$\Omega_{\rm m}$ plane, these plots illustrate the comprehensiveness of our proposed emulator, as it can emulate all three cosmologies perfectly without any retraining. Finally, our emulator also does not output $\sigma_8$. Therefore, we post-process the emulator chains in \textsc{CoCoA} to compute $\sigma_8$ without recalculating the likelihood values or altering the weights of the accepted points in the MCMC. This process is also trivially parallelizable.}
    \label{fig:posterior_shifts}
\end{figure}

Fig.~\ref{fig:posteriors} compares all cosmological parameter posteriors predicted by \textsc{CoCoA} against the \textsc{ResTRF256} emulator trained on the enhanced set. In this chain, the data vector was computed in \textsc{CoCoA} at the fiducial cosmology defined in Table~\ref{tab:cosmology}. This basic verification illustrated that the excellent \textsc{ResTRF512}  $\Delta \chi^2$ accuracy translates into a superb agreement at the posterior level. We then test if the emulator retains its accuracy when the cosmology is shifted with respect to the fiducial cosmology. This involves shifting the fiducial cosmology along the first principal component in the $\sigma_8$-$\Omega_{\rm m}$ plane using standard normalization, given by
\begin{equation}
\label{eq:pc}
    \mathrm{PC}=0.012\,\Omega_{\rm m} + 0.0096\,\sigma_8\,,
\end{equation}
as shown in Fig.~\ref{fig:posterior_shifts}. 

These shifts in the principal component result in relatively moderate changes in the $\Omega_{\rm m}$-$\log(10^{10} A_{\rm s})$ plane. Nonetheless, the contours in Fig.~\ref{fig:posterior_shifts} illustrate the volume's comprehensiveness in parameter space well is emulated by our proposed neural network. Notably, the agreement between \textsc{CoCoA}  and \textsc{ResTRF512} at the edges of the $\Omega_{\rm m}$-$\log(10^{10} A_{\rm s})$ two-dimensional posterior distribution on the second and third cosmologies is particularly encouraging. The modeling of these tails is not trivial; the projected spatial $\Delta \chi^2$  distribution presented in Fig.~\ref{fig:val_diff_temperatures} shows a scarcity of training points when $\Omega_{\rm m}$ and $\log(10^{10} A_{\rm s})$ are simultaneously low or high even when $T=512$. Filling these regions with additional training points will require adjustments in our training strategy that we will explore in future work.

\subsection{Third Validation Level: Bayesian Evidence}

The third level of validation involves comparing \textsc{CoCoA} and the emulators in terms of their ability to compute Bayesian evidence. Bayes' theorem, which underpins the foundation of Markov Chain Monte Carlo (MCMC) methods, relates conditional probabilities within data and parameter spaces as follows:
\begin{equation}
    \mathcal{P}(\boldsymbol \theta) = \frac{\Pi(\boldsymbol \theta)\mathcal{L}(\boldsymbol \theta)}{\mathcal{Z}}\,.
\end{equation}
In this equation, $\mathcal{P}(\boldsymbol \theta)$ is the posterior distribution, $\Pi(\boldsymbol \theta)$ the prior, $\mathcal{L}(\boldsymbol \theta)$ the likelihood, and $\boldsymbol \theta$ is the vector of cosmological and nuisance parameters. The normalization constant, $\mathcal{Z}$, is the evidence, which quantifies the probability of observing the data given a model. Several competing samplers have been developed to compute the evidence, displaying varying degrees of accuracy and robustness. Following the comparative analysis in~\citet{Miranda:2020lpk,2023MNRAS.521.1184L}, we adopted the \textsc{PolyChord} sampler and set the hyperparameters to the conservative values \textsc{precision criterion} $= 0.001$, $n_\mathrm{live} = 1024$, and $n_{\rm repeats}$ equals to five times the number of sampled dimensions~\citep{2015MNRAS.450L..61H,2015MNRAS.453.4384H}. 

Table~\ref{tab:evidence} summarizes our comparison across the same three cosmologies shown in Fig.~\ref{fig:posterior_shifts}. The comparison suggests that the logarithm of Bayes factor, $\log \mathcal{Z}$, is tolerant to errors on the order of $\Delta \chi^2 \sim \mathcal{O}(1)$. Specifically, the differences between \textsc{CoCoA} and \textsc{ResMLP512} are $\Delta  \log \mathcal{Z} =(0.39 \pm  0.23, 0.39 \pm  0.23, 0.84 \pm  0.22)$ For cosmologies $1$, $2$, and $3$, respectively. The more precise \textsc{ResTRF256} emulator reduces these differences to $\Delta \log \mathcal{Z} = (-0.14 \pm 0.22, 0.05 \pm 0.23,0.5 \pm 0.22)$. Also, \textsc{ResTRF128} predicted $\Delta \log \mathcal{Z} = (-0.07 \pm 0.23,0.26 \pm 0.23, -0.34 \pm 0.22)$. In contrast, the more comprehensive \textsc{ResTRF512} predicted $\Delta \log \mathcal{Z} = (-0.16 \pm 0.22, 0.09 \pm 0.23, -0.35 \pm 0.22)$, indicating that parameter space coverage, at this temperature range, has a negligible effect on the evidence and the error in $\log\mathcal{Z}$ corresponds to the emulation error.

Generally, the accuracy target for $\log \mathcal{Z}$ is $\sigma(\log_{10} \mathcal{Z}) \lesssim 1/\sqrt{3}$, so the uncertainty on Bayesian evidence ratio $R \equiv \mathcal{Z}_{\rm AB}/(\mathcal{Z}_{\rm A}\mathcal{Z}_{\rm B})$, a method that assesses whether an assumed model can explain the data measured by two independent experiments 
$A$ and $B$ with a single set of parameters ($AB$ refers to their joint likelihood), is $\sigma(\log_{10} (R)) \lesssim 1$. These bounds are inspired by the Jeffreys scale, which states that $ \log_{10} (R) > 2.3$ reflects strong support for the hypothesis that the model can simultaneously explain datasets $A$ and $B$, while $ \log_{10} (R) < -2.3$ indicates strong tension~\citep{Jeffreys:1939, DES:2020hen}. Both \textsc{ResMLP} and \textsc{ResTRF} seem suitable to compute Bayesian evidence, and the lower $T = 128$ temperature can accommodate moderate parameter shifts with sufficient accuracy for assessing tension. However, the \textsc{ResTRF256} and  \textsc{ResTRF512} emulators offer both accuracy and better coverage in parameter space.

\subsubsection{Sampler Comparison: \textsc{Polychord} vs. \textsc{Nautilus}}

All emulators significantly speed up the evidence calculation compared to \textsc{CoCoA}. In particular, the \textsc{ResTRF}-\textsc{PolyChord} combination consumed approximately 100 CPU hours to compute $\log(\mathcal{Z})$ on the conservative hyperparameter settings. The runtime was a bit over one hour when using one \textsc{Intel Sapphire Rapids Xeon  96-core CPU} node, a resource still in the exclusive domain of supercomputers. Nonetheless, we can combine our emulators with the ongoing efforts from the community to develop faster samplers to compute evidence. For example, we compared the promising \textsc{Nautilus}~\citep{lange2023nautilus} sampler against \textsc{Polychord} results. We ran the \textsc{Nautilus}, using the \textsc{ResTRF256} to compute the data vectors, with the hyperparameter $n_\mathrm{live}=4000$ and got $\log\mathcal{Z} = (-20.38, -21.61, -18.96)$. In this case, the evaluation takes about one hour and utilizes only one core. Finally, Bayesian evidence can be computed quickly on portable computers.

\begin{table}
    \centering
    \renewcommand{\arraystretch}{1.1}
    \begin{tabular}{|l|c |c|c|c|} \hline
        & $N_\mathrm{train} $ & Cosmo 1     & Cosmo 2     & Cosmo 3    \\\hline
        \textsc{CoCoA}    	  & ---                     & $-20.21 $ 	  & $-21.71$	   & $-18.56$     \\ 
         			    	   &	                     & $\pm 0.16$  & $ \pm 0.16$ & $\pm 0.15$ \\ \hline \hline
	    \textsc{ResTRF256}  & $3 \times 10^6$	         &  $-20.35$ 	  & $-21.66$ 	  & $-19.06 $   \\
						      & (enhanced set) &  $\pm 0.15$ 	& $\pm 0.16$ 	& $\pm 0.15$ \\ \hline \hline
        \textsc{ResMLP512}	&$3 \times 10^6$	& $-20.6 $ 	  & $-22.1 $ 	& $-19.4$    \\ 
         				    &		& $\pm 0.16$  & $\pm 0.16$ 	& $\pm 0.15$ \\  \hline
        \textsc{ResTRF512}	& $3 \times 10^6$	& $-20.37$ 	    & $-21.62$ 		& $-18.91 $  \\
	 				          &		  & $\pm 0.15$ 	  & $\pm 0.16$ 	  & $\pm 0.15$ \\ \hline \hline
	    \textsc{ResTRF512}  & $1.2 \times 10^6$	           &  $-20.40$ 	    & $-21.36$ 	    & $-19.18 $    \\
						    & (superior set)   &  $\pm 0.15$ 	& $\pm 0.16$    & $\pm 0.15$  \\ \hline
	    \textsc{ResTRF512}  & $6 \times 10^5$	           &  $-20.59$ 	    & $-21.76$ 	    & $-18.97 $    \\
						    &  (standard set)  & $\pm 0.16$ 	& $\pm 0.16$ 	& $\pm 0.15$  \\ \hline \hline						
	    \textsc{ResTRF128}   	& $3 \times 10^6$	& $-20.28$ 		& $-21.45$ 		& $-18.90 $   \\
						          &       & $\pm 0.16$ 	  & $\pm 0.16$ 	  & $\pm 0.15$ \\ \hline 
    \end{tabular}
    \caption{Emulators validation against \textsc{CoCoA} on the computation of the logarithm of the Bayesian evidence, $\log\mathcal{Z}$. In all cases, we adopted the \textsc{PolyChord} nested sampler with the conservative hyperparameters \textsc{precision criterion} $= 0.001$, $n_\mathrm{live} = 1024$ and $n_{\rm repeats}$ equal $5$ times the number of dimensions sampled~\citep{2015MNRAS.450L..61H}. The trailing integer in the emulator name is the training temperature, and $N_{\rm train}$ is the number of training points. The three cosmologies differ by moderate shifts in the $(\log(10^{10} A_{\rm s})$-$\Omega_{\rm m})$ plane (see Fig.~\ref{fig:posterior_shifts}).}
    \label{tab:evidence}
\end{table}

\section{Tension Metrics}

Quantifying the discordance of datasets in cosmology is a non-trivial task due to non-Gaussianities in posteriors. This is seen predominantly on marginalized posteriors, where the non-Gaussianities can be hidden in the marginalized dimension. Additionally, since the parameter space can be of large dimension for some cosmological models, methods relying on Bayesian evidence or direct integration can be non-tractable. As such, considerable effort has been put forth to test different metrics for quantifying tension~\citep{raveri_concordance_2019,raveri_non-gaussian_2021,park_concordance_2020,DES:2020hen}.

An additional challenge is posed by the computation time for the chains used to calibrate the metrics. Every metric needs a series of chains at fixed shifts from a fiducial cosmology. We generate $256$ noise realizations on the data by randomly sampling from the LSST-Y1 cosmic shear likelihood. The noise realizations are generated separately for each shifted cosmology, and further shift the cosmological parameters in directions that are not known a priori. With the standard \textsc{CoCoA} pipeline, each MCMC would take about $250$ CPU-hours, and each \textsc{PolyChord} run about $2 \times 10^4$ CPU-hours on a \textsc{Intel Sapphire Rapids Xeon  96-core CPU} node. 

Instead, we apply the \textsc{ResTRF}$512$ emulator trained on the enhanced set to demonstrate how this process can be significantly accelerated. Using our emulator, each MCMC takes about $20$ CPU-hours, while each \textsc{PolyChord} run takes about $100$ CPU-hours on the same hardware. The inclusion of evidence-based metrics is particularly intractable with the standard pipeline. We use this to show how noise realizations of the data can affect the tension metrics, and which tension metrics are most robust to noise realizations of the data. 

To examine the effects of emulation error on the tension metrics, we compare the tension metrics using chains generated with several of the emulators tested in the previous section. Although the emulation error is generally not enough to alter the conclusion drawn from the tension metrics, there is a residual effect. These are more pronounced with the \textsc{ResMLP} emulator, which had the highest emulation error. These effects do not generally persist across all metrics, however. We demonstrate that the emulation error has a negligible impact on the reported tension.

\subsection{Defining Tension Metrics}

Tension metrics are a method of quantifying the discordance between two datasets. Due to non-Gaussianities in posterior distributions, tension metrics and must be calibrated for each pair of experiments. This calibration can be done by injecting a shift on exactly one parameter and comparing the metrics to the Gaussian error. Previously, the Dark Energy Survey collaboration has calibrated tension metrics between its data and Planck 2018~\citep{DES:2020hen}.  The DESC collaboration ought to redo such studies in the context of Rubin Observatory to better interpret the results of the many Bayesian tools DESC intends to adopt when analyzing their upcoming year one data. The authors in~\citet{DES:2020hen} shift the cosmology by up to $5\sigma$ in both $\sigma_8$ and $\Omega_{\rm m}$ directions. They then run a single chain and compute the tension metrics in the $\boldsymbol \theta = (\log(10^{10} A_{\rm s}),n_{\rm s},H_0,\Omega_{\rm b},\Omega_{\rm m})$ parameter space. 

This manuscript contrasts~\citet{DES:2020hen} by running additional chains at multiple noise realizations on the data. We generate a set of $256$ data vectors randomly sampled from the likelihood for each shift. The tension is computed at each noise realization, allowing us to understand the robustness of tension metrics against noise. To accelerate generating chains, we employ the \textsc{ResTRF} emulator trained on the enhanced set to compute the LSST-Y1 cosmic shear data vectors, and \textsc{Cosmopower} to emulate the CMB power spectra from Planck. The CMB likelihood used in \textsc{Cosmopower} is a reimplementation of the Planck2018 lite high-$\ell$ TTTEEE likelihood~\citep{Prince:2019hse,Planck:2019nip,2022MNRAS.511.1771S}.

Many tension metrics have been developed to compare two datasets, which we label $A$ and $B$.~\citep{raveri_concordance_2019}. Depending on the data required to compute the tension, these come in different classes. Parameter-space metrics only involve the posterior $\mathcal{P}(\boldsymbol{\theta})$, the prior $\Pi(\boldsymbol{\theta})$, or the likelihood $\mathcal{L}(\boldsymbol{\theta})$. Contrasting this approach are the evidence-based metrics, which involve statistics relating to Bayes' factor $\mathcal{Z}$. There are also hybrid methods, such as the goodness-of-fit degradation, which rely on the posterior, prior, and likelihood but not on an MCMC chain. For all metrics, we can define a probability to exceed (PTE), and the tension strength $N_\sigma$ as the number of Gaussian standard deviations~\citep{raveri_non-gaussian_2021}
\begin{equation}\label{eq:n_sigma}
    N_\sigma = \sqrt{2}\mathrm{Erf}^{-1}({\rm PTE})\,.
\end{equation}

To generate chains, we use the \textsc{Emcee} ensemble sampler~\citep{Foreman_Mackey_2013} with 120 walkers to ensure our chains have more than enough samples. The Gelman-Rubin diagnostic is not applicable to the ensemble sampler due to the constant communication among the walkers. Instead, we allowed each walker to run at least $35$ times the estimated autocorrelation length ($\tau_{\rm ACL}$), and $\tau_{\rm ACL} \sim 750$ in our chains. In the Appendix, we demonstrate that this convergence is strong enough for tension calibration.

\subsubsection{Parameter Difference}

        The parameter difference is a powerful metric as it does not require datasets to be uncorrelated, nor does it require Gaussianity in the posterior or the likelihood~\citep{PhysRevD.101.103527, raveri_non-gaussian_2021, DES:2020hen}. Suppose we have samples from posteriors $\mathcal{P}(\boldsymbol{\theta}_{\rm A})$ and $\mathcal{P}(\boldsymbol{\theta}_{\rm B})$, and define the parameter difference as $\Delta\boldsymbol{\theta} = \boldsymbol{\theta}_{\rm A} - \boldsymbol{\theta}_{\rm B}$. Under this reparameterization, the two posteriors are $\mathcal{P}(\boldsymbol{\theta}_{\rm A})$ and $\mathcal{P}(\boldsymbol{\theta}_{\rm A}-\Delta\boldsymbol{\theta})$. 
        
        Whenever the posteriors are independent, one can marginalize over $\boldsymbol{\theta}_{\rm A}$ to get the parameter difference distribution:
        \begin{equation}
            \mathcal{P}(\Delta\boldsymbol{\theta}) = \int \mathcal{P}(\boldsymbol{\theta}_{\rm A})\mathcal{P}(\boldsymbol{\theta}_{\rm A} - \Delta\boldsymbol{\theta}) \,d\boldsymbol{\theta}_{\rm A}
        \end{equation}
        In the complete absence of tension between the chains $\boldsymbol{\theta}_{\rm A}$ and $\boldsymbol{\theta}_{\rm B}$, $\mathcal{P}(\Delta\boldsymbol{\theta})\leq\mathcal{P}(0)$ for all $\Delta\boldsymbol{\theta}$ as they are centered around the same point $\boldsymbol\mu$. Therefore, PTE can be defined as the volume of the posterior contours with $\mathcal{P}(\Delta\boldsymbol{\theta})>\mathcal{P}(0)$ as follows.
        \begin{equation}
            \mathrm{PTE} = \int\limits_{\mathcal{P}(\Delta\boldsymbol{\theta})>\mathcal{P}(0)}\mathcal{P}(\Delta\boldsymbol{\theta}) \, d\Delta\boldsymbol{\theta}
        \end{equation}
        
        To compute the tension using parameter difference, we employ the method of normalizing flows~\citep{raveri_non-gaussian_2021,papamakarios_masked_2018,germain_made_2015}. The flow is constructed using neural networks that learn a diffeomorphism to map samples between two probability densities. In practice, the target distribution to map to is a normal distribution. Suppose the mapping is given by a diffeomorphism $F$ that maps points following a normal distribution $\mathcal{N}(0,1)$ to points following the parameter difference distribution $\mathcal{P}(\Delta\boldsymbol{\theta})$. The probability density of the parameter difference distribution can then be found using the probability density of a normal distribution and the Jacobian of $F$,
        \begin{equation}
            P(\Delta\boldsymbol{\theta}) \approx F^{-1}(\Delta\boldsymbol{\theta})|\det J(F)|\,.
        \end{equation}
        
        To determine $F$, we use a Masked Autoregressive Flow (MAF) constructed using a sequence of Masked Autoencoders for Density Estimation (MADE) in \textsc{TensorFlow}. Each MADE performs a transformation on the input that follows the autoregressive property,
        \begin{equation}
            y_i = T(x_{1:i-1})+S(x_{1:i-1})x_i\,,
        \end{equation}
        with each $x_{i:i-1}$ corresponding to a masked input,
        \begin{equation}
            x_{1:i-1} = (x_1,\ldots,x_{i-1},0,\ldots,0)\,.
        \end{equation} 
        The functions $T(x_{i:i-1})$ and $S(x_{i:i-1})$ are neural networks containing learnable weights. To train the normalizing flow, we use a batch size of $2048$, a validation split of $0.1$, a learning rate of $10^{-2}$, and we train for $100$ epochs. For the remaining hyperparameters, we follow~\citet{raveri_non-gaussian_2021}, where the hidden dimension and the number of MADEs is $2$ times the number of parameters, which is $5$ for this application. An examination of the convergence of the normalizing flow is performed in Appendix~\ref{appendix_a}.
        
        In addition, we follow \cite{raveri_non-gaussian_2021} and implement a `pre-whitened' parameter space which is related to the $\Lambda$CDM parameter space by the linear transformation
        \begin{equation}
            \boldsymbol{\theta}' = \mathcal{C}^{-1}(\boldsymbol{\theta}-\boldsymbol\mu)\,,
        \end{equation}
        with $\mathcal{C}$ the covariance and $\boldsymbol\mu$ the mean of the cosmological parameters $\boldsymbol{\theta}$. The results between the original parameter space and `pre-whitened' space are consistent; however, pre-whitening the parameter space improves the neural network's convergence rate.

\subsubsection{Parameter difference in update form}

        Proposed in~\citet{raveri_concordance_2019}, this metric looks at how the posterior changes when adding in a second data set. Then we define a parameter $Q_{\mathrm{UDM}}$ by looking at the difference between chain $A$ and chain $AB$ as
        \begin{equation}
            Q_{\mathrm{UDM}} = (\boldsymbol \mu_{\rm A} - \boldsymbol\mu_{\rm AB})^T (\mathcal{C}_{\rm A}-\mathcal{C}_{\rm AB})^{-1}(\boldsymbol\mu_{\rm A} - \boldsymbol\mu_{\rm AB})
        \end{equation}
        If the posteriors are Gaussian, then $Q_{\mathrm{UDM}}$ is $\chi^2$ distributed with $d=\text{rank}(\mathcal{C}_{\rm A}-\mathcal{C}_{\rm AB})$ degrees of freedom, representing the number of parameters which become more constrained when adding a second data set. Thus, we can define the PTE as
        \begin{equation}
            \mathrm{PTE} = \int\limits^{Q_{\mathrm{UDM}}}_{0}\chi^2(x,d) \, dx \,.
        \end{equation}
        Here, $\chi^2(x,d)$ is the $\chi^2$ distribution with $d$ degrees of freedom, and $x$ is an integration variable.
        
        Computations of $Q_{\mathrm{UDM}}$ can be noisy, resulting in tensions which are nonsensical when compared to the a priori tension. To alleviate this, we follow~\citet{raveri_concordance_2019} and~\citet{DES:2020hen} by performing a Karhunen-Lo\'eve (KL) mode. This amounts to solving for the generalized eigenvalues of $\mathcal{C}_{\rm A}$ weighted by $\mathcal{C}_{\rm AB}$,
        \begin{equation}
            (\mathcal{C}_{\rm A} -  \lambda \mathcal{C}_{\rm AB})\boldsymbol{\theta} = 0
        \end{equation}
        We filter out the noisy contributions by restricting the calculations to modes with 
        \begin{equation}
            1.05 < \lambda^i < 100\,,
        \end{equation}
        where the $\lambda^i$ are the weighted eigenvalues of the KL decomposition. The lower and upper bounds filter out the KL modes that are not updated when adding the other dataset~\citep{raveri_concordance_2019}. 

        This procedure requires Gaussianity of the posterior. This assumption is violated by LSST-Y1 cosmic shear, where unconstrained parameters are approximately uniformly distributed. However, with this metric, there is a choice of which dataset corresponds to $A$. Since \textsc{Cosmopower} and joint likelihood chains are nearly Gaussian in all parameters, we can use these two to compute $Q_\mathrm{UDM}$. 

\subsubsection{Goodness-of-fit Degradation}

        This is another metric described in~\citet{raveri_concordance_2019}, and it examines how the goodness-of-fit changes when adding a second data set. If the experiments have Gaussian likelihoods, we can compute $\chi^2$ for each chain $\boldsymbol{\theta}_{\rm A}$ and $\boldsymbol{\theta}_{\rm B}$ and compare it to $\chi^2$ of the joint chain $\boldsymbol{\theta}_{\rm AB}$ as
        \begin{equation}
            Q_{\mathrm{DMAP}} = 2\mathcal{L}_{\rm A}(\hat{\boldsymbol{\theta}}_{\rm A}) + 2\mathcal{L}_{\rm B}(\hat{\boldsymbol{\theta}}_{\rm B}) - 2\mathcal{L}_{\rm AB}(\hat{\boldsymbol{\theta}}_{\rm AB})\,,
        \end{equation}
        where $\hat{\boldsymbol{\theta}}_{\rm A}$ is the maximum a posteriori of the dataset $A$. $Q_{\mathrm{DMAP}}$ is $\chi^2$ distributed with degrees of freedom $d=N-\mathrm{tr}(C^{-1}_\Pi C_\mathcal{P})$ with $N$ the number of parameters. The ratio of the variance of the posterior to the variance in the prior estimates the number of constrained parameters by the likelihood/data. Thus, the PTE is
        \begin{equation}
            \mathrm{PTE} = \int\limits_{0}^{Q_{\mathrm{DMAP}}}\chi^2(x,d) \, dx\,,
        \end{equation}

        This procedure requires Gaussianity in both the cosmological parameters and data space. For LSST-Y1, the data space is Gaussian, but the parameter space is not. For the CMB chains, the parameter space is approximately Gaussian.

        We use the \textsc{py-bobyqa} minimizer to find the maximum a posteriori and the \textsc{ResTRF} emulator to compute the likelihood. We start each optimizer at the mean of the posterior, allowing for a smaller initial region of trust for the optimizer. When using our emulator, each optimizer run takes $\mathcal{O}(1)$ minute.
        
\subsubsection{Bayesian Suspiciousness}
        
        Bayesian evidence acts as a normalization constant in Bayes' theorem. One can use the ratio of evidence to approximate the agreement between datasets:
        \begin{equation}
            \log R = \log\mathcal{Z}_{\rm AB}-\log\mathcal{Z}_{\rm A}-\log\mathcal{Z}_{\rm B}\,.
        \end{equation}
        Because each $\mathcal{Z}$ requires integration over the entire parameter space supported by the prior, the Bayesian evidence naturally depends on the volume of the prior~\citet{DES:2020hen, Handley_2019_tension}. To account for this, one can introduce a quantity called the \textit{information} derived from the Kullback-Leibler divergence $D$, a number quantifying the amount of information gained from the likelihood~\citep{kl_div, Handley_2019_bayes}. The information is given by
        \begin{equation}
            \log I = D_{\rm A}+D_{\rm B}-D_{\rm AB}\,.
        \end{equation}
        By taking the difference between $\log R$ and $\log I$, the prior dependence is removed, and the remaining part is the tension from the datasets alone. The difference is called the \textit{suspiciousness} given by
        \begin{equation}
            \log S = \log R - \log I\,.
        \end{equation}
        
        To compute the PTE, one can again determine the number of dimensions constrained by the likelihood by~\citep{Handley_2019_bayes}
        \begin{equation}
            d = d_G - 2\log S\,,
        \end{equation}
        where $d_G$ is the Bayesian model dimensionality. This number is $\chi^2$ distributed with $d_G$ degrees of freedom. 
        
        Thus, the probability to exceed is given by
        \begin{equation}
            \mathrm{PTE} = \int\limits_d^\infty \chi^2(x,d_G)\,dx\,.
        \end{equation}
        We use the \textsc{Anesthetic}~\citep{anesthetic} package to compute the suspiciousness and Bayesian model dimensionality from the \textsc{PolyChord} output. Computing these additional statistics comes with a small amount of noise, which results in variations of $N_\sigma$ on the order of $0.1$ for cosmologies $1$, $4$, and $5$. Since we do not consider the noise from \textsc{Anesthetic} when evaluating tension metrics, we relax the criteria for detecting a bias in the tension metrics when using different emulators.

\subsubsection{Eigentension}

        First proposed by~\citet{park_concordance_2020}, this metric aims to remove poorly measured eigenvectors of the covariance where the tension is dominated by the prior rather than the likelihood. The steps to compute are as follows:
      	\begin{enumerate}
            \item Find the eigenvalues and eigenvectors of chain A covariance.
            \item Find the ratio of the variance in the prior and the posterior. Consider this mode `well-measured' if the ratio is greater than $10^2$.
            \item Project chain B onto the well-measured eigenvectors of A.
            \item Compute the parameter difference PTE only using the well-measured eigenmodes. 
        \end{enumerate}
        In practice, chain $A$ will be the LSST-Y1 chain, as it has some unconstrained parameters that will cause the eigentension to differ from the other metrics. We compute the tension on the two well-measured eigenmodes of the LSST-Y1 chain using the parameter difference method described above. Following this procedure, there is no assumption of Gaussianity in the posterior or the likelihood.

\begin{table}
    \centering
    \setlength{\tabcolsep}{1.2pt}
    \renewcommand{\arraystretch}{1.4}
    \begin{tabular}{|l|c |c|c|c|c|c|} \hline
          & $ N_\mathrm{train}$  & $\Delta N_\sigma^{\rm PD}$     & $\Delta N_\sigma^{\rm E}$    & $\Delta N_\sigma^{\rm QUDM}$  &  $\Delta N_\sigma^{\rm QDMAP}$ &  $\Delta N_\sigma^{\rm S}$     \\\hline
        \textsc{ResTRF256}   	& $3 \times 10^6$	&  $0$ & $0$ & $0$ & $0$ & $0$ \\ \hline\hline
        \textsc{ResTRF512}		& $3 \times 10^6$	&  $-0.03$ & $-0.06$ & $0.01$ & $-0.01$ & $-0.23$ \\ \hline
        \textsc{ResTRF512}   	& $1.2 \times 10^6$	&  $-0.07$ & $0.04$ & $0.01$ & $0.00$ & $-0.26$ \\ \hline
        \textsc{ResTRF512}   	& $6 \times 10^5$	&  $-0.04$ & $0.06$ & $-0.17$ & $0.00$ & $-0.13$ \\ \hline\hline					
        \textsc{ResMLP512}		& $3 \times 10^6$	& $0.07$ & $0.15$ & $0.06$ & $0.09$ & $-0.20$ \\ \hline
        \textsc{ResMLP512}		& $1.2 \times 10^6$	& $0.04$ & $0.19$ & $-0.06$ & $-0.06$ & $3.04$ \\ \hline
    \end{tabular}
    \caption{Validation of how the various tension metrics are affected by the differences in the $\Delta \chi^2$ performance between the multiple \textsc{ResTRF512} and \textsc{ResMLP512} emulators. The exceptional \textsc{ResTRF256} precision allows its results to be considered the ground truth. The tension metrics were computed between the Planck data and the LSST synthetic cosmic shear data centered at the fiducial cosmology shown in Table~\ref{tab:cosmology}. The superscript PD corresponds to {\it Parameter Difference}, E to {\it Eigentension}, QUDM and QDMAP to $Q_{\rm UDM}$ and $Q_{\rm DMAP}$ metrics, and S to {\it Bayesian Suspiciousness}. We have not added noise to the LSST fiducial data vector. We see that, in all cases using the \textsc{ResTRF}, the tension metrics are well below the $|\Delta N_\sigma^X|<0.3$ threshold. However, the \text{ResMLP512} trained on the superior set fails this threshold of $|\Delta N_\sigma^X|<0.4$ for suspiciousness. This reflects the loss of accuracy in $\log\mathcal{Z}$ when using \textsc{ResMLP}.} 
    \label{tab:tension_fiducial}
\end{table}

\begin{table}
    \centering
    \setlength{\tabcolsep}{1.2pt}
    \renewcommand{\arraystretch}{1.4}
    \begin{tabular}{|l|c |c|c|c|c|c|} \hline
         	&  Cosmo & $\Delta N_\sigma^{\rm PD}$    & $\Delta N_\sigma^{\rm E}$    & $\Delta N_\sigma^{\rm QUDM}$  &  $\Delta N_\sigma^{\rm QDMAP}$ &  $\Delta N_\sigma^{\rm S}$     \\\hline
         \textsc{ResTRF512}		& 4 &  $0.06$ & $0.02$ & $0.01$ & $0.00$ & $-0.04$ \\ \hline
         \textsc{ResTRF256}		& 4	&  $0$    & $0$    & $0$     & $0$    & $0$     \\ \hline
         \textsc{ResTRF128}   	& 4	&  $0.02$ & $0.02$ & $-0.01$ & $0.00$ & $-0.05$ \\ \hline\hline					
         \textsc{ResTRF512}		& 5	&  $-0.10$ & $0.10$ & $0.04$ & $-0.01$ & $-0.17$ \\ \hline
         \textsc{ResTRF256}		& 5	&  $0$     & $0$    & $0$     & $0$     & $0$     \\ \hline
         \textsc{ResTRF128}   	& 5	&  $-0.21$ & $0.07$ & $0.04$ & $0.00$  & $-0.12$ \\ \hline	
    \end{tabular}
    \caption{Validation of how the various tension metrics are affected by the difference in training coverage between the multiple \textsc{ResTRF} emulators; they were all trained on the enhanced set that contains $N_\mathrm{train} = 3 \times 10^6$ models. Although there are accuracy differences between them, we expect any lack of coverage to have a far more significant effect on the results of these metrics, which were computed between the Planck data and the LSST synthetic cosmic shear data centered at two cosmologies, Cosmo 4 and Cosmo 5. Their exact parameter values are shown in Table~\ref{tab:cosmology}. The superscript PD corresponds to {\it Parameter Difference}, E to {\it Eigentension}, QUDM and QDMAP to $Q_{\rm UDM}$ and $Q_{\rm DMAP}$ metrics, and S to {\it Bayesian Suspiciousness}. We have not added noise to the LSST data vectors. For all of the temperatures considered, $\Delta N_\sigma^X$ is less than our threshold of $0.3$ at the shifted cosmologies. This indicates the results will not depend on the choice of temperature when $T\geq128$.} 
    \label{tab:tension_shifts}
\end{table}

\section{Propagating Emulator Errors on Tension Metrics}

Despite the promising posterior-level accuracy of the \textsc{ResTRF} emulator, we still need to check whether the emulation errors bias the tension metrics. This check is particularly important for the parameter difference and suspiciousness metrics, which require data in the tails of the posterior near the prior boundaries. We test how the \textsc{ResTRF} and \textsc{ResMLP} emulators trained with different temperatures and a different number of training points affect the resulting tension. To quantify this, we compute the absolute value of the difference in $N_\sigma$ between a given emulator and the \textsc{ResTRF}$256$ trained on the enhanced set, which we consider truth due to its high accuracy. We consider a shift of $|\Delta N_\sigma^X | = 0.3$ to be the threshold where the bias becomes significant for the parameter-based metrics and $|\Delta N_\sigma^S| = 0.4$ for suspiciousness. The results at the cosmology $1$, both for \textsc{ResMLP} and \textsc{ResTRF}, are summarized in Table~\ref{tab:tension_fiducial}. 

We only find a significant bias in the results for the suspiciousness metric when using the \textsc{ResMLP512} emulator trained on the superior set. This can be explained by the noticeable improvement in $\langle \Delta \chi^2 \rangle$ from the superior set to the enhanced set. The results of the Eigentension metric are not considered significant. However, they stand out from the others with a $\Delta N_\sigma^{\mathrm{E}}\sim0.2$. The difference is likely sourced by slight parameter biases in the constrained parameters from the \textsc{ResMLP}, such as $\log(10^{10}A_s)$, which are reflected in the eigenmodes. The other metrics do not have significant biases for any of the emulators tested, even for parameter difference, which relies on sampling within the tails of the posterior. 

In Table~\ref{tab:tension_shifts}, we examine the bias of tension metrics at the shifted cosmologies we consider when evaluating tension metrics. In this case, we find no significant bias in any of the metrics at any of the shifts for any of the emulators. The most significant bias, however, came from the \textsc{ResTRF128} emulator at cosmology $5$, with a shift of $\Delta N_\sigma^\mathrm{PD}=-0.21$. This is likely a compounding effect of the reduced parameter space covered by the emulator and the priors, making it difficult to generate sampling points in the region of lower $\Omega_\mathrm{m}$ and larger $\log{(10^{10}A_{\rm s})}$. Nevertheless, the bias is not significant enough to alter the interpretation of the tension.

The results of Tables~\ref{tab:tension_fiducial} and~\ref{tab:tension_shifts} indicate it is safe to use any of the \textsc{ResTRF} emulators to calibrate the tension metrics. This reflects the marginal effect the loss of accuracy has at the posterior and the Bayesian evidence level. Since we do not examine the best-fit cosmology of each noise realization, we opt to use the \textsc{ResTRF512} emulator trained on the enhanced set to ensure noise realizations are contained within the training data.

\begin{figure}
    \centering
    \includegraphics[width=0.87\columnwidth]{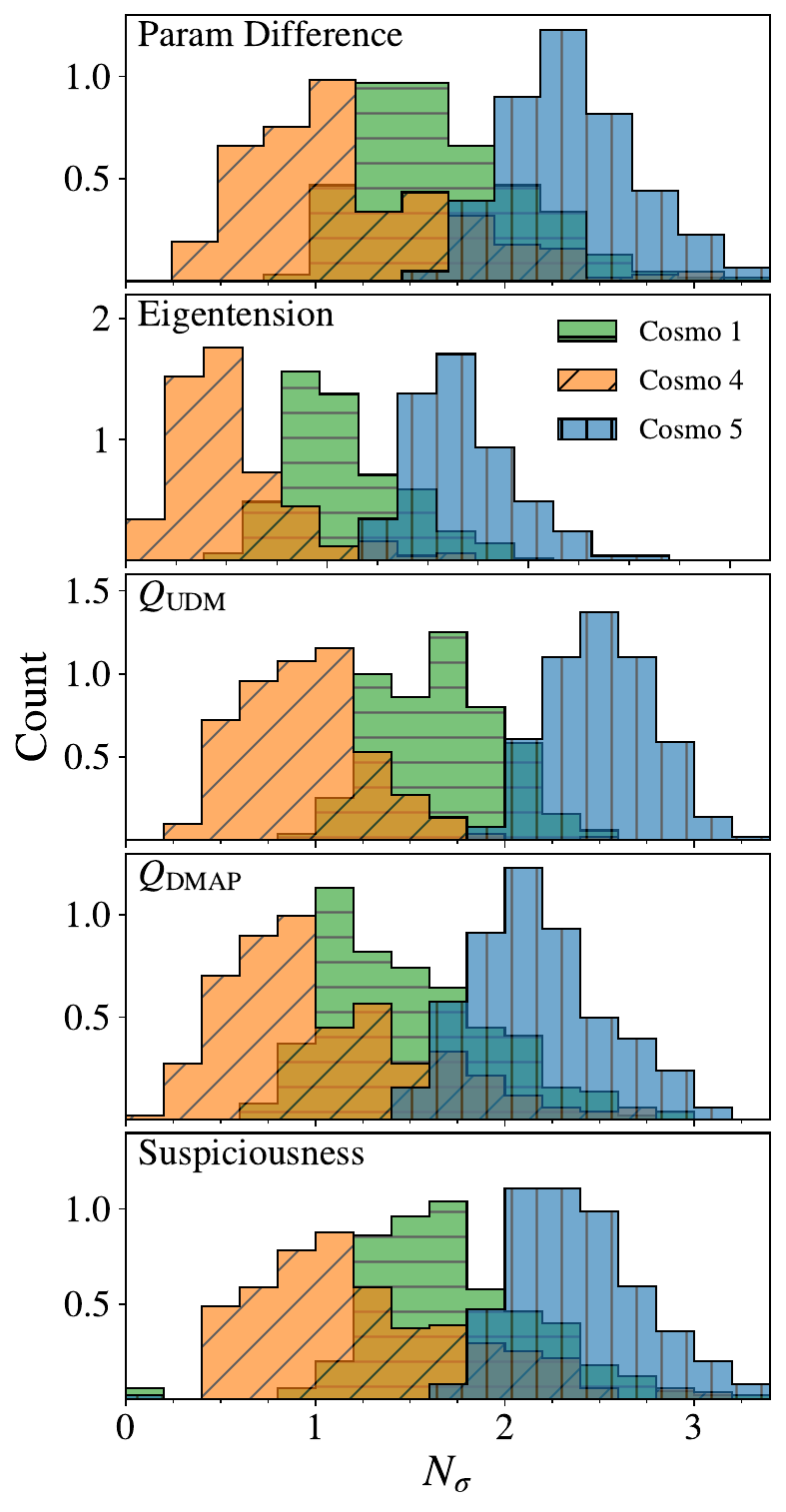}
    \caption{\textit{From top to bottom }: Parameter difference, eigentension, parameter difference in update form, goodness-of-fit degradation, and Bayesian suspiciousness. Cosmology 1 is the fiducial cosmology given in Table~\ref{tab:cosmology}, while Cosmology 4 and 5 are shifted in the $\Omega_{\rm m}$-$\sigma_8$ plane and are defined in Table~\ref{tab:cosmology}. The posteriors are highly non-Gaussian and, as such, do not exactly agree with the a priori tension. We find that the eigentension consistently reports a much lower tension than any of the other metrics. It does, however, have the narrowest distribution of $N_\sigma$, suggesting it is robust against noise realizations of the data. The remaining four tension metrics generally agree across noise realizations. They all tend to give higher tension estimates than the a priori estimate. This is particularly pronounced in the parameter difference in update form, which reports a larger tension than all other metrics when the injected tension is largest.}
    \label{fig:tension_metrics}
\end{figure}

\section{Tension results}

\subsection{A Priori Tension}

The tension metrics are calibrated using cosmology 1 (Table~\ref{tab:cosmology}), and cosmologies 4 and 5 (Table~\ref{tab:cosmology}). The injected tension at each cosmology can be estimated by computing the $\chi^2$ between LSST and Planck,
\begin{equation}
    \chi^2 = \delta\boldsymbol{\theta}^T (\mathcal{C}_{\mathrm{LSST}} + \mathcal{C}_{\mathrm{Planck}})^{-1} \delta\boldsymbol{\theta}\,.
\end{equation}
Here, $\delta\boldsymbol{\theta}$ is the two-dimensional vector describing the difference in $\log(10^{10} A_{\rm s})$ and $\Omega_{\rm m}$ between the LSST and Planck chains, and $\mathcal{C}_{(\cdot)}$ is the $2\times2$ covariance matrix of the respective chain in the $\log(10^{10} A_{\rm s})$-$\Omega_{\rm m}$ plane. This quantity follows a $\chi^2$ distribution with $2$ degrees of freedom, from which the probability to exceed and $N_\sigma$ can be computed. In summary, we find the a priori tension with no shift to be $N_\sigma = 1.5$, and the a priori tension to be $N_\sigma = 1.1$ and $N_\sigma = 1.8$ for cosmology $4$ and $5$, respectively.

The a priori tension makes use of a Gaussian approximation of the posteriors in the $\log(10^{10} A_{\rm s})$-$\Omega_{\rm m}$ plane, and is a good estimate of the tension that was injected by shifting the cosmologies. As such, we use this as a comparison point for each of the described tension metrics. However, the a priori tension is not expected to match the result of each tension metric. Firstly, the noise realizations further shift the cosmological parameters. Secondly, any non-Gaussianity is ignored.

\subsection{Results}

A summary of the results of each tension metric is provided in Fig.~\ref{fig:tension_metrics}. Aside from the eigentension, each metric agrees, on average, with the a priori tension estimate when the tension is lowest at cosmology 4. As the injected tension increases, the a priori tension tends to underestimate the tension reported by each of the metrics. In contrast, the eigentension has the reverse effect: it tends to have a more substantial agreement with the a priori estimate at cosmology 5, which has the largest deviation from Planck. 

Comparing the distribution of the tension metrics to each other, the distributions of the parameter difference and the suspiciousness match most closely, with the suspiciousness varying more significantly with noise realizations. These metrics make the least amount of assumptions about the underlying data space and parameter space distributions and don't require cutting any data. These metrics, however, have stochastic noise that increases the variation compared to the other metrics. The parameter difference metric has noise from the neural network, while the suspiciousness has noise in the computation of the Bayesian model dimensionality $d_G$ and Kullback-Leibler divergence $D$. We empirically find that these deviations can generate variations in $N_\sigma$ of order $0.1$. 

Eigentension, which removes the unconstrained directions from the posterior, significantly reduces the tension in the $\Omega_\mathrm{m}$-$\log(10^{10} A_{\rm s})$ plane while also giving tighter distribution with respect to the noise realizations, despite the noise from the normalizing flow still being present. However, removing hard-prior effects in the posterior may make training the normalizing flow more stable. Interestingly, this discrepancy in $N_\sigma$ is not present when computing the tension in the $\log(10^{10} A_{\rm s})$-$\Omega_{\rm c} h^2$.

The $Q_{\mathrm{UDM}}$ and $Q_\mathrm{DMAP}$ metrics have Gaussianity requirements: the former in parameter space and the latter in both parameter and data spaces. Despite this, both metrics strongly agree with the parameter difference and suspiciousness, but with much tighter variation. The $Q_{\mathrm{UDM}}$ metric gives slightly higher $N_\sigma$ than the parameter difference at each noise realization. In contrast, the agreement between $Q_{\mathrm{DMAP}}$ and parameter difference is great; they agree well at each noise realization as well. However, the tighter distribution demonstrates that this metric is more robust against noise realizations on the data than the other metrics. This is likely reflective of the frequentist approach taken to $Q_{\mathrm{DMAP}}$, depending primarily on the best-fit cosmology rather than the posterior.

As mentioned above, this procedure for calibrating tension metrics required the evaluation of $768$ LSST-Y1 cosmic shear with our emulator and $768$ joint likelihood chains with our emulator and \textsc{CosmoPower}. Since each chain is run at both shifts in parameter space and data space, we demonstrate that our emulator can accurately calibrate tension metrics without retraining an emulator at each shift. Additionally, our emulator remains accurate when adding the \textsc{CosmoPower} reimplementation of the Planck CMB 2018 lite high-$\ell$ TTTEEE likelihood is used, which can further shift the parameters~\citep{Prince:2019hse,Planck:2019nip,2022MNRAS.511.1771S}.

\section{Conclusion}

The growing precision of cosmological analyses presents a computational challenge. In this paper, we focus on the real-space cosmic shear correlation function for LSST-Y1 and present a neural network emulator based on the transformer architecture~\citep{vaswani_attention_2017} that generalizes the analysis presented in ~\citet{zhong2024attentionbased}.  We extend the pure transformer architecture by combining it with \textsc{ResMLP} blocks and allowing more freedom in the weights at each position of the transformer. Furthermore, we further test a procedure for generating training data that generalizes better to higher dimensions and larger priors and training datasets. Overall, the scaled \textsc{ResTRF} emulator speeds up likelihood evaluation times by about three orders of magnitude over \textsc{CoCoA}, while using fewer computational resources. 

We generate our training data using a Gaussian approximation on the cosmological and nuisance parameters using a covariance derived from an MCMC. We then define the temperature $T$ as a scaling factor for the covariance that expands the volume covered by the Gaussian approximation. By doing this, we can cover a significant volume of parameter space without worrying about the curse of dimensionality, which becomes especially important when considering models that require new parameters beyond $\Lambda$CDM. Additionally, this allows us to use more precise codes to generate the data vectors with minimal sacrifice to the training time. Once the cosmological parameters are known, data vector computation is trivially parallelizable, allowing us to use a larger training set.

By using $\langle \Delta \chi^2 \rangle$ as our loss function, we have a natural metric to assess the accuracy of our emulator. Only the \textsc{ResTRF} architecture can have most testing models with $\langle \Delta \chi^2 \rangle \lesssim 0.2$ at the highest temperature we tested, $T=512$. We also test the emulator at the posterior level, where there is the complete agreement between the emulator and \textsc{CoCoA}. Additionally, we shift the cosmology along the most constraining direction by $\pm20\sigma$ and demonstrate that our emulator still gives robust posteriors as well as good estimates of the Bayesian evidence. The \textsc{ResTRF} emulator outperformed the \textsc{ResMLP} in the accuracy of the Bayesian evidence and is accurate enough that it can be combined with other efforts to accelerate the sampling for Bayesian evidence, such as \textsc{Nautilus}. 

Since we rely on the average $\Delta \chi^2$ as our loss function, the training can be degraded by outliers that drive up the average. By simply removing the outliers, we can greatly improve the accuracy of our emulator. Outlier mitigation will be explored in a subsequent work (part III of this series). Additionally, we must have an estimate of the parameter covariance to generate our training samples. This can become difficult in cases where an MCMC is intractable. In part III, we will explore other ways to generate the training data that do not rely on an MCMC, such as doing Fisher or DALI approximation. 

Our emulator is then applied to calibrating tension metrics, a procedure that requires numerous chains at various cosmologies. We extend the typical tension calibration by running chains centered at cosmologies represented by noise realizations of the data. These further shift the cosmologies and require us to run hundreds of chains. With our emulator, this process is tractable. Including evidence-based metrics means we must run hundreds or chains using \textsc{PolyChord}; however, our emulator also makes this exercise tractable. 

Using our analysis, we find considerable agreement between all tension metrics aside from the eigentension. More work must be done to determine the exact cause, which is beyond the scope of this manuscript. Meanwhile, the goodness-of-fit degradation gave the most tightly distributed tension across the noise realizations, despite the violation of Gaussianity assumptions by the LSST-Y1 cosmic shear chains. The heightened variation seen in parameter difference, parameter difference in update form, and suspiciousness could be due to stochastic noise. This effect could be reduced by averaging over several evaluations of the tension. 

\section*{Acknowledgements}

We thank Eduardo Rozo and Marco Bonici for their valuable discussions and careful reading of the manuscript. TE is supported by the Department of Energy HEP-AI program grant DE-SC0023892. Simulations in this paper use High-Performance Computing (HPC) resources supported by Stony Brook Research Computing and Cyberinfrastructure, and the Institute for Advanced Computational Science at Stony Brook University for access to the high-performance SeaWulf computing system, which was made possible by a $\$1.4$M National Science Foundation grant ($\#1531492$).

\bibliography{tension_paper} 


\appendix*

\section{Normalizing Flow Convergence}\label{appendix_a}

\begin{figure}
    \centering
    \includegraphics[width=0.95\columnwidth]{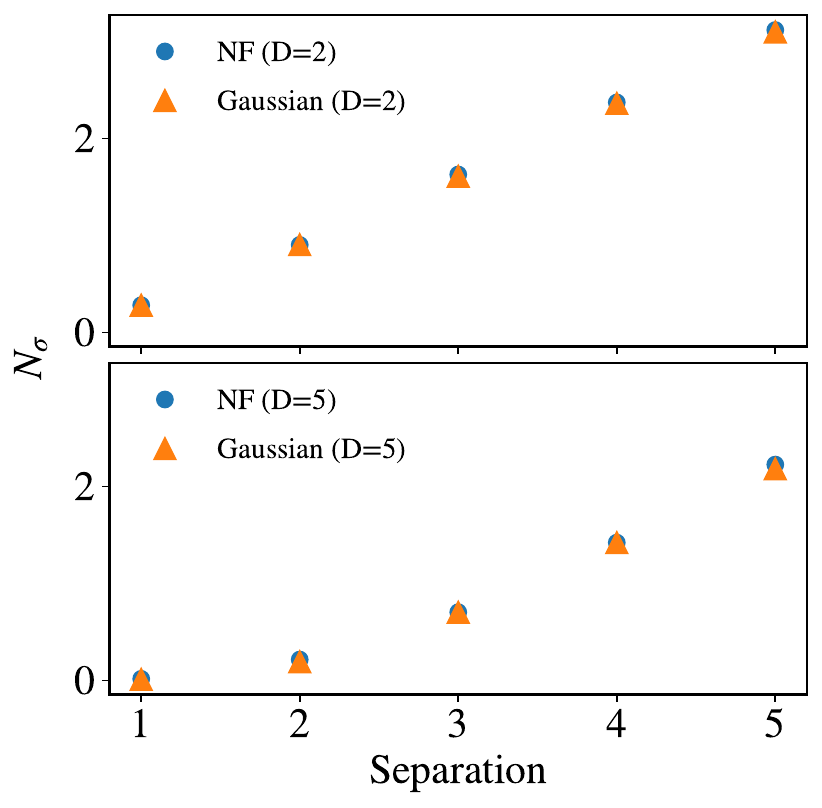}
    \includegraphics[width=\columnwidth]{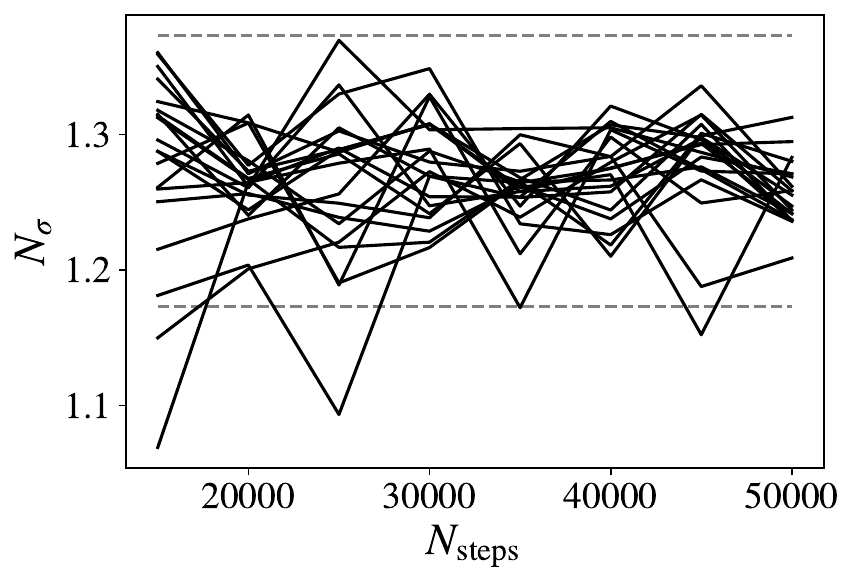}
    \caption{
    \textit{Top:} We see that, in both two and five dimensions, the tensions agree at separations between $1$ and $5$. To achieve this, only $5 \times 10^4$ points are required to achieve convergence. In this manuscript, our chains consist of $2.4 \times 10^5$ points. As such, we have shown that our implementation of normalizing flows converges and that the chains we used consist of enough points to ensure an accurate mapping is learned. 
    \textit{Bottom:} We plot the result of the normalizing flow result as a function of the number of steps of each MCMC walker. For every case, we take $5 \times 10^3$ steps for burn-in and compute the tension for every additional $5 \times 10^3$ steps, starting with $1.5 \times 10^4$ steps. At each chain length, we run $20$ normalizing flows on identical samples. We do not include noise realizations on the data. We find that the noise in the tension result is dominated by noise in the normalizing flow rather than the length of the chain. 
    }
    \label{fig:ten_v_n}
\end{figure}

Normalizing flows (NFs) are neural networks that learn a diffeomorphism that shifts samples to be Gaussian distributed. As such, it is important to demonstrate that normalizing flows converge to the correct solutions in cases where a solution can be obtained analytically. To complete this, we sample $5 \times 10^4$ points from a multivariate Gaussian distribution with unit covariance and mean at the origin. The distribution is then shifted by $1$ to $5$ along a single axis, where $5 \times 10^4$ more points are generated at each shift. We compute the Gaussian tension and the tension using NF between the shifted and centered at the origin samples using Eq.~(\ref{eq:n_sigma}). For each shift, we then compare the results. This procedure is done in both two and five dimensions. 

As displayed in Fig.~\ref{fig:ten_v_n}, we find complete agreement at all shifts tested. This indicates that the neural network parameterizing the normalizing flow is giving consistent results when there are $5 \times 10^4$ samples. Even though the transformation is trivial, the neural network does not know that; hence, it is reasonable to assume the neural network can learn mappings for moderately non-Gaussian distributions. To ensure this is true, we generate $2.4 \times 10^5$ samples for each chain, far exceeding the $5 \times 10^4$ used in this test.

Lastly, we check the normalizing flow computation as a function of the number of steps of each MCMC walker. We run our chains using the \textsc{emcee} sampler with $120$ walkers and take the first $5000$ steps as burn-in. We then compute the tension using normalizing flows for every $5000$ steps, starting from $15000$ steps. The computation is repeated $20$ times, giving an estimate of the error of the normalizing flow. For this test, we do not include any noise realizations on the data; all computations are done at cosmology $0$, defined in Table~\ref{tab:cosmology}. In Fig.~\ref{fig:ten_v_n}, we find that running the chain for more than $25000$ steps is sufficient to keep the error in $N_\sigma$ within $0.1$. Additionally, we see that there is no bias in the results for chains with more than $20000$ steps. 

\label{lastpage}
\end{document}